\documentclass[twocolumn,showpacs,superscriptaddress,10pt]{revtex4-2}
\usepackage{amsmath,amssymb,graphicx, subfigure, hyperref, braket, float}
\usepackage{blindtext}

\counterwithout{figure}{section}
\usepackage[export]{adjustbox}
\graphicspath{{figure/}}
\usepackage[T1]{fontenc}
\usepackage{ae,aecompl}
\usepackage{graphicx, xcolor}
\begin{document}

\title{Distribution of distance-based quantum resources outside a radiating Schwarzschild black hole}

\author{Samira Elghaayda}
\email{samira.elghaayda-etu@etu.univh2c.ma}
\address{Laboratory of High Energy Physics and Condensed Matter, Department of Physics, Faculty of Sciences Ain Chock, Hassan II University, P.O. Box 5366 Maarif, Casablanca 20100, Morocco.}
\author{Xiang Zhou}
\email{202010106112@mail.scut.edu.cn}
\address{Department of Mathematics, South China University of Technology,\\ Guangzhou, 510640, People’s Republic of China}
\author{Mostafa Mansour}
\email{mostafa.mansour.fpb@gmail.com}
\address{Laboratory of High Energy Physics and Condensed Matter, Department of Physics, Faculty of Sciences Ain Chock, Hassan II University, P.O. Box 5366 Maarif, Casablanca 20100, Morocco.}

\begin{abstract}
We obtain analytical expressions for distance-based quantum resources and examine their distribution in the proximity of a Schwarzschild black hole (SBH) within a curved background. For an observer in free fall and their stationary counterpart sharing the Gisin state, the quantum resources are degraded at infinite Hawking temperature. The extent of this degradation that occurs as the SBH evaporates is contingent upon the fermionic frequency mode, Gisin state parameters, and the distance between the observer and the event horizon (EH). In the case of two accelerating detectors in Minkowski spacetime interactiong with quantum fluctuating scalar fields (QFSF), we find that quantum coherence and discord exhibit sudden disappearance for certain initial states and sudden reappearance for others except entanglement, regardless of the Unruh temperature. We also discover that the quantum resources of one detector can be transferred to another in the case of two stationary detectors through a fluctuating quantum field outside the SBH. We demonstrate that, in contrast to coherence and discord, we are unable to regenerate entanglement for a given initial state and that they are equal for different vacuum states. In certain circumstances, the presence of EHs does not significantly reduce the available resources, as it turns out that all interesting phenomena occur within EHs. Since the world is basically non-inertial, it is necessary to understand the distribution of quantum resources within a relativistic framework.

\end{abstract}


\maketitle

\section{Introduction}
Entanglement \cite{horodecki29} is a valuable resource for quantum-based technologies \cite{furusawa1, gisin2, pant2}, and has attracted significant interest \cite{mansour2021bipartite, elghaayd1, yu2005,alsing2006}. Furthermore, theoretically, entanglement may not fully describe all quantum correlations in a quantum system \cite{henders, vedra,modi}. According to \cite{ollivier2}, these quantum correlations are well-known as quantum discord. It is remarkable that for a separable mixed state, quantum discord can remain nonzero. While entanglement remains a primary resource for many quantum tasks, quantum discord has also been recognized for its operational significance and its ability to serve as a resource for mixed-state quantum computing \cite{passante2}, even in the absence of entanglement. Quantum coherence, resulting from the superposition principle \cite{streltsov2, orszag20}, is even more fundamental than entanglement \cite{strelts}. Similar to entanglement, quantum coherence can be employed as a resource in various quantum tasks \cite{winte, chitambarg}. Essentially, quantum coherence, discord, and entanglement each characterize the quantum nature of a system from different perspectives. Therefore, to effectively utilize these resources, it is essential to appropriately quantify them across multiple systems.

In recent decades, quantum field theory (QFT) has gained significant attention for exploring quantum resources and their application in various protocols \cite{summers1m, summers1987bell, lin2015quantum, lapponi2}. Within QFT, particle detector models have emerged as a powerful tool for addressing diverse challenges \cite{sorkin19e, benincasa2,perche2022localized}. These models offer an operationally attractive approach to QFT phenomenology in curved spacetimes, encompassing phenomena like the Hawking and Unruh effects \cite{hawking19,unruh19, takagi1986vacuum, hodgkinson2014static}. The Hawking effect posits that QFT in the spacetime of a SBH leads to the emission of Hawking radiation \cite{hawking19}. Later, Unruh made a significant prediction based on the Hawking effect, stating that an observer in flat Minkowski spacetime would detect a vacuum state, while another observer hovering near the EH of SBH with uniform acceleration would detect it as a thermal state \cite{unruh19, crispino2008unruh,elghaayda2023entropy}. The temperature detected by an accelerating particle detector is proportional to its acceleration $T_U=\frac{\hbar a_u}{2\pi c k_B}$, where light travels at the speed $c$, the reduced Planck constant is $\hbar$, and $k_B$ is the Boltzmann constant.
\begin{figure}[H]
	\centering
	\includegraphics[width=0.7\linewidth]{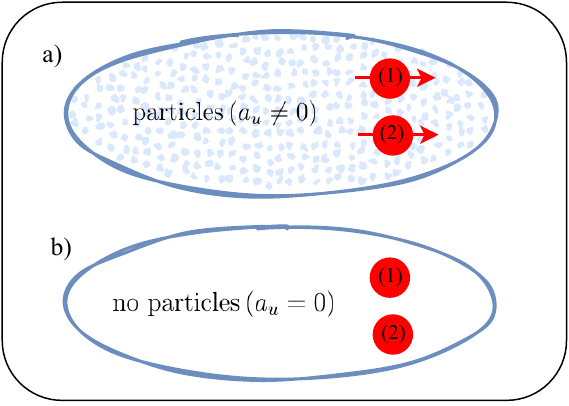}
	\caption{Diagram $a)$ represents the case of accelerated detectors, while $b)$ represents the stationary detectors.}
	\label{fig:c}
\end{figure}
As seen in Fig. \ref{fig:c}, an accelerating detector will detect particles in an empty flat space, but a stationary detector would not detect particles in an empty flat space. Furthermore, for curved space times, such as those used to characterize black holes, the problem becomes more complex; a follow-up explanation in this regard will be provided for different vacuum states later. Beyond their utility as indispensable tools, particle detector models are used in the fields of relativistic quantum optics and information \cite{martin2020general, martin2013processing}. 

Recently, considerable attention has been devoted to exploring the implications of the Unruh and Hawking effects on quantum resources \cite{hu2011entanglement,martin201,alsing2006,benatti2004entanglement}. It has been proven that both entanglement \cite{fuentes2005alice} and teleportation fidelity \cite{alsing200} are found to be sensitive to the Unruh and Hawking effects, suggesting that the concept of a vacuum relies on the observer's trajectory through space-time \cite{hu2011entanglement,benatti2004entanglement}. 
Subsequently, it was confirmed that different field types impact the degradation of Unruh-induced entanglement in qualitatively different ways. In particular, a pair of moving atoms with constant velocity interacts with a QFSF without mass in the Minkowski vacuum prepared in a separable state turn out to be entangled as a manifestation of the Unruh effect \cite{benatti2004entanglement}. Additionally, it has been observed that the Hawking effect-induced entanglement degradation raises the entropic uncertainty bound \cite{feng2015uncertainty}. Since the distribution of quantum resources in a relativistic framework has been further investigated in relativistic quantum information, this is still an open question worth exploring.

As a further step along this line, we will examine the distribution of distance-based quantum resources  between two observers, ($O_1$) and ($O_2$) sharing a Gisin state for Dirac fields in the background of a SBH. In this case, ($O_1$) is falling freely toward a SBH near its horizon, while ($O_2$) remains at a small distance from the EH SBH. We then consider two detectors, which are independent of each other and moving with constant velocity outside a radiating SBH. To analyze this situation, we treat it as an open system that interacts with a fluctuating bath of coupled QFSF without mass. However, when dealing with a vacuum state in curved spacetime, a challenge arises in determining the vacuum state of quantum fields. Therefore, in this paper, we focus on the Boulware and Hartle-Hawking (HH) vacuum states.

In this article, we begin by examining the fundamental concepts used in this investigation to quantify the quantum resources, as outlined in Sect. \ref{sec2}. These concepts include Helinger distance coherence, trace distance discord, and Bures distance entanglement. We perform these quantifications within the Schwarzschild spacetime (SST) framework and between two particle detectors in curved spacetime. To facilitate this analysis, we extensively review the vacuum structure and Hawking effect for Dirac particles in SST in Sect. \ref{sec0}. In addition, we study two particle detector systems using an exactly solvable Hamiltonian and derive the master equation that governs their evolution in Sect. \ref{sec1}. We explore their dynamics in the HH and Boulware vacuum states. In Sec. \ref{sec4}, we describe the key findings gained throughout this investigation. We summarize and discuss our conclusions in the last Sec. \ref{sec5}. As a convention, we use the unit system $k_B=\hbar=c=1$ in this paper.
\section{Distance-based measurements of quantum resources \label{sec2}}
For the purposes of this paper, it is useful to review the concepts of Helinger distance coherence, trace distance discord, and Bures distance entanglement, which are used as quantum resource quantifiers.
\subsection{X states}
Through numerical demonstration, it was shown that all bipartite mixed states can be reduced to X states with a single entanglement-preserving unitary transformation; as a result, the concurrence and other entanglement measures of the resulting X state are equal to those of the former general state. This corresponds to a density matrix that is provided by
\begin{equation}\label{xs}
	\mathcal{S}=\left(
	\begin{array}{cccc}
		s_{11} & 0 & 0 & s_{14} \\
		0 & s_{22} &s_{23}  & 0 \\
		0 & s_{23} & s_{33} & 0 \\
		s_{14} & 0 & 0 & s_{44}
	\end{array}
	\right).
\end{equation}
In this case, $s_{14}$ and $s_{23}$ are real, and $s_{11}+s_{22}+s_{33}+s_{44}=1$. Non-negativity requirements $s_{14}^2\leqslant s_{11}s_{44}$ and $s_{23}^2\leqslant s_{22}s_{33}$. It is generally agreed upon that X-states are useful in a variety of experimental settings \cite{bos,rau2}. Numerous well-known state classes are included in this class, including isotropic states \cite{yu2005}, Werner states \cite{wernm}, and diagonal Bell states.. X-states have also been studied in a variety of quantum information fields. Especially within the relativistic framework of quantum information, as we will trait in this work.
\subsection{Coherence based on Helinger distance} 
Quantum coherence is an essential resource for quantum information tasks \cite{winte, chitambarg}. Several measures of coherent quantum systems have been discussed in literature, for instance, the l$_1$-norm, relative entropy, skew information, and Hilenger distance \cite{baumgratz2, hu201, streltsov20, jin}. In particular, as shown in Ref. \cite{herbut2005} for a finite-dimensional abstract space $N$, they fix an orthogonal basis $\left\lbrace\left| j\right\rangle_{k=1,...,N} \right\rbrace $ representing an observable $K$. This is because their focus is on coherence from a basis-dependent perspective. Obviously, coherence was obtained by an appropriate application of the Hellinger distance between the two density matrices, $\mathcal{S}_{1}$ and $\mathcal{S}_{2}$ \cite{roga2016}, which is
\begin{equation}
	d_H(\mathcal{S}_{1},\mathcal{S}_{2})=\sqrt{tr\left[ (\sqrt{\mathcal{S}_{1}}-\sqrt{\mathcal{S}_{2}})^2\right] },
\end{equation}
where the definition of the square root is understood throughout this work to be
\begin{equation}
	\left\langle k\left| \sqrt{\mathcal{S}_{1}}\right| k\right\rangle =\sqrt{\left\langle k\left| \mathcal{S}_{1}\right| k\right\rangle},
\end{equation}
which deviates slightly from the standard definition of a matrix's square root. A comparable occurrence of square roots can be noted in classical-optic coherence issues \cite{martinez1984, martinez1986r}. Thus, we define the coherence quantifier as the distance to the nearest incoherent state $\mathcal{S}_{p}$, based on the Hellinger distance \cite{jin2018}.
\begin{equation}
	\mathcal{C}_H=\left[ d_H(\mathcal{S},\mathcal{S}_{p})\right]^2=tr\left[ (\sqrt{\mathcal{S}}-\sqrt{\mathcal{S}_{p}})^2\right],
\end{equation}
With incoherent we indicate states diagonal in the reference basis, so that $\mathcal{S}_{p}$ is the diagonal component of $\mathcal{S}$ in the same basis.
\begin{equation}
	\mathcal{S}_{p}=\sum_{k=1}^{N}s_{k,k}\left| k\right\rangle \left\langle k\right|,
\end{equation}
where $s_{i,k}=\left\langle i\left| \mathcal{S}\right| k\right\rangle $ are the matrix elements of $\mathcal{S}$ in the basis $\left\lbrace \left| k\right\rangle \right\rbrace $. This measure can be written like this:
\begin{equation}\label{c1}
	\mathcal{C}_H=\sum_{k\neq j}\left|s_{k,j} \right|.
\end{equation}
The underlying assumption of this definition is that the base-dependent non-diagonal terms of a state's density matrix are essentially what determines its coherence on a given basis.
\subsection{Discord based on trace distance}
Evaluating quantum discord analytically is a challenging task \cite{ollivier2}. As a matter of fact, a geometric measurement of quantum discord was introduced \cite{tufarel, dakic2}. This measurement quantifies quantum correlations by calculating the minimal Hilbert-Schmidt distance from classical states. This method has certain advantages, as the procedure of minimization can be carried out theoretically for arbitrary bipartite states. However, one of its unfavorable properties is that it can increase with local operations performed in an unquantified component \cite{piani20}. Indeed, this defect can be addressed by employing another norm in the set of states. The optimal option is to utilize the Schatten 1-norm to define quantum discord \cite{paula2013}. The trace distance discord for a bipartite system $\mathcal{S}$ is defined as the minimum trace distance between $\mathcal{S}$ and all classical-quantum states $\mathcal{S}_{cq}$ \cite{paula2013}, specifically
\begin{eqnarray}\label{DDT}
	\mathcal{D}_{\rm T}= \frac{1}{2} \min_{\chi_\mu\in\mathcal{S}_{cq}}||\mathcal{S}-\varSigma||_1,
\end{eqnarray}
In this case, $||\mathcal{S}-\varSigma||_1={\rm Tr}\sqrt{(\mathcal{S}-\varSigma)^\dag (\mathcal{S}-\varSigma)}$ defines the trace norm. the $\mathcal{D}_{\rm T}$ measures the distance between the investigated quantum state $\mathcal{S}$ and any classical-quantum state $\varSigma$ in the collection $\mathcal{S}_{cq}$ of classical-quantum states. The formulation of a general classical-quantum state $\varSigma\in\mathcal{S}_{cq}$ is $\mathcal{S}_{cq} = \sum_k p_k ~P_{k,1}\otimes \mathcal{S}_{k,2}$, in which the state of the second component is $\mathcal{S}_{k,2}$, the set of corresponding probabilities is denoted by $\{p_k\}$, the orthogonal projectors acting on the first component is represented by $P_{k,1}$, and the set of corresponding probabilities is $\{p_k\}$. Additionally, the $\mathcal{D}_{\rm T}$ is only given for bipartite X states \cite{paula2013, ciccarello2014}.
\begin{eqnarray}\label{d1}
	\mathcal{D}_{\rm T}= \sqrt{\frac{\Upsilon_{1}^2
			\Upsilon_{\rm max}^2-\Upsilon_{2}^2 \Upsilon_{\rm min}^2}
		{\Upsilon_{\rm max}^2-\Upsilon_{\rm min}^2+\Upsilon_{1}^2-\Upsilon_{2}^2}},
\end{eqnarray}
where $\Upsilon_{1,2}=2(|s_{23}|\pm|s_{14}|)$,  $\Upsilon_{3}=1-2(s_{22}+s_{33})$, $\Upsilon_{\rm min}^2=\min\{\Upsilon_{1}^2,\Upsilon_{3}^2\}$, $\Upsilon_{\rm max}^2={\rm \max} (\Upsilon_{3}^2,\Upsilon_{2}^2+\Upsilon_{30}^2)$, $\Upsilon_{03}=2(|s_{11}|+|s_{44}|)-1$ and $\Upsilon_{30}=2(|s_{11}|+|s_{22}|)-1.$

\subsection{Entanglement based on Bures distance}
The Bures distance \cite{bu01,bu02,bu04} is one of the best approaches to assessing the inseparability of a quantum state \cite{ma1, ma2, elghaayda2ntum}. It is calculated by using the geometric distance between the state and the nearest separable state in the system; for a bipartite $\mathcal{S}$ state, this distance is derived geometrically by
\begin{equation}\label{bures}
	\mathcal{B}_d= \sqrt{2 -2\sqrt{\mathcal{F}_d}},
\end{equation}
where, as determined by the fidelity, $\mathcal{F}_d$ denotes the level of separability of $\mathcal{S}$. The mathematical function $\mathcal{F}_d$ is
\begin{equation}
	\mathcal{F}_d= \max_{\sigma\in S}F\left(\mathcal{S},\sigma\right)=\frac{1}{2}\left(1+\sqrt{1-\mathcal{C}^{2}}\right)\label{eq:Fs-2},
\end{equation}
where all separable states ${ \sigma \in S }$ are taken into consideration to establish the maximum value of Uhlmann's fidelity $F\left(\mathcal{S},\sigma\right)$ \cite{bu02}. The bipartite X-state (\ref{fpr}) concurrence is provided by
\begin{eqnarray}
	\label{lg8}
	\mathcal{C}=2\max\{ 0, \lambda_1,\lambda_2 \}.
\end{eqnarray}
where $\lambda_1=s_{14}-\sqrt{s_{22} s_{33}}$ and $\lambda_2=s_{23}-\sqrt{s_{11} s_{44}}$. For maximally entangled states, the $\mathcal{B}_d$, has a greatest value of $\mathcal{B}_d=(2-\sqrt{2})^{1/2}$ and for separable states is $\mathcal{B}_d=0$. We employ a normalized version of this entanglement metric
\begin{equation}\label{bures01}
	\mathcal{B}_d=\frac{1}{\sqrt{2-\sqrt{2}}} [\sqrt{2 -\sqrt{2+ 2\sqrt{1-\mathcal{C}^2}}} ],
\end{equation}
So, for maximally entangled states, $\mathcal{B}_d=1$, and for separable states, $\mathcal{B}_d=0$.
\section{Radiating SBH \label{sec0}}
This section deals with the effect of Hawking radiation on scalar particles and the structure of the vacuum of spacetime as perceived by observers of a bipartite Gisin state, which is maximally entangled for the observer freely falling into a SBH.
\subsection{Vacuum structure and Dirac fields}
\begin{figure}[H]
	\centering
	\includegraphics[width=1\linewidth]{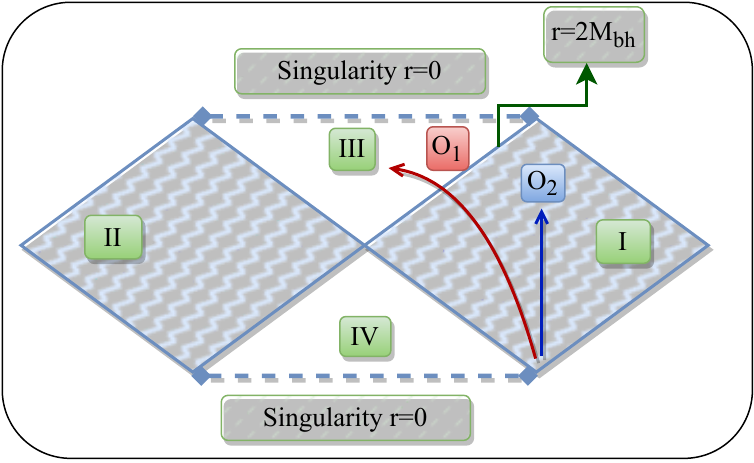}
	\caption{A Penrose diagram is used to represent SST and showing trajectories for ($O_1$) and ($O_2$), where ($O_1$) is free-falling toward the SBH and ($O_2$) is static. Region III represents the inside of the SBH while regions I and II represent the outside of the SBH. The white hole is located in region IV.}
	\label{fig:b}
\end{figure}

To examine the distribution of quantum resources in the presence of a SBH, we first need to revisit the notion of vacuum states in curved spacetime. Specifically, we will focus on a SBH, where the line element for the SST can be described by
\begin{equation}\label{be}
	ds_{bh}^2=-(1-\frac{2M_{bh}}{r} ) dt'^2+(1-\frac{2M_{bh}}{r})^{-1} dr^2+r^2 d\Omega_{bh}^2,
\end{equation}
where the line element on the unit sphere is $d\Omega_{bh}^2$ and the mass of the SBH is $M_{bh}$. As seen in Fig. \ref{fig:b}, the observer outside the SBH is unable to acquire information about the particle states inside the horizon because of the existence of an EH at $R_{bh} = \frac{1}{2M_{bh}}$. The vacuum experienced by a stationary observer is thermal, as described by thermo-field dynamics \cite{israel1976}, and has a Hawking temperature of $T_H=\frac{\varsigma}{2\pi}$, with a surface gravity of $\varsigma=\frac{1}{4M_{bh}}$.
For Dirac field witout mass that fulfills $[i\gamma^y(\partial_y -\Gamma_y)]\psi_D = 0$, where $\Gamma_y$ is the spin connection. Kruskal coordinates specify its vacuum structure for different observers
\begin{equation}
	\mathcal{U}=-4M_{bh}e^{-\varsigma(t'-r^*)}, \quad \mathcal{V}=4M_{bh}e^{\varsigma(t'+r^*)},
\end{equation}
where the tortoise coordinate is $r^* = r + 2M_{bh}\ln\left| 1 - \frac{r}{2M_{bh}}\right| $. As a function of the newly established coordinates, Eq (\ref{be})'s highest mathematical extension is
\begin{equation}
	ds_{bh}^2=-\frac{1}{2\varsigma r}e^{-2\varsigma r}d\mathcal{U}d\mathcal{V}+r^2(d\xi^2+\sin^2\xi d\zeta^2),
\end{equation}
The Rindler manifold approximates the near-horizon structure, allowing us to examine the vacuum structure of a related quantum field to that of an accelerated observer in flat space \cite{pan2008,wang2010}.The HH vacuum $\left| 0_H\right\rangle$ is equivalent to the Minkowski vacuum for ($O_1$) in a free fall with a proper timelike vector $\partial_{t'}\varpropto (\partial_\mathcal{U}+\partial_\mathcal{V})$. A static ($O_2$) with timelike Killing vector, $\partial_{t'}\varpropto (\mathcal{U}\partial_\mathcal{U}-\mathcal{V}\partial_\mathcal{V})$, is analogous to the Rindler vacuum in flat space in terms of the Boulware vacuum $\left| 0\right\rangle_I$ corresponding to positive frequencies associated with $\partial_t$. Another vacuum $\left| 0\right\rangle_{II}$ associated with $-\partial_t$ was also detected.  

Despite this, caution must be exercised regarding the reliability of the analogy mentioned above, as it only applies in the vicinity of the SBH. The proper acceleration for ($O_2$) located at a position $d_0$ close to the EH is given by $a_{bh}=\frac{\varsigma}{\sqrt{1-2M_{bh}/d_0}}$. This establishes a constraint that the Rindler approximation is meaningful only if $\frac{\Delta_H}{R_{bh}}\leqslant1$ where $\Delta_H=d_0-R_{bh}$ \cite{martin201}. Considering that $\left| 0_H\right\rangle\equiv\otimes_j\left| 0_{\omega_j}\right\rangle_H $, and its first excitation $\left| 1_H\right\rangle\equiv\otimes_j\left| 1_{\omega_j}\right\rangle_H $, we can write the HH vacuum and its excitation in the Boulware basis as \cite{martin201, alsing2006}
\begin{equation}\label{sh}
		\begin{array}{cc}
  \begin{aligned}
			\left| 0_{\omega_j}\right\rangle_H& =\frac{1}{\sqrt{1+e^{-\vartheta}}} \left| 0_{\omega_j}\right\rangle_I\left| 0_{\omega_j}\right\rangle_{II}  \\
			& + \frac{1}{\sqrt{1+e^{\vartheta}}} \left| 1_{\omega_j}\right\rangle_I\left| 1_{\omega_j}\right\rangle_{II}\\
			\left| 1_{\omega_j}\right\rangle_H& =\left| 1_{\omega_j}\right\rangle_I\left| 0_{\omega_j}\right\rangle_{II}
   \end{aligned}
		\end{array},
\end{equation}
where $\varpi=2\pi \omega/ \varsigma =8\pi \omega M_{bh}$ is the mode frequency determined by ($O_2$) with regard to of the surface gravity, $R_0=d_0/R_{bh}=d_0/2M_{bh}$, and $\vartheta=\varpi\sqrt{1-1/R_0}$. Notably, apart from its dependency just on the Hawking temperature, we can also examine the quantum resources as a function of the distance of ($O_2$) to the EH through the parameterization in (\ref{sh}) as demonstrated in \cite{pan2008, wang2010}.
\subsection{Gisin state in SST}
An attractive set of states known as Gisin states were introduced in \cite{gisin}, which are defined as combinations of pure, intricate, and some separable mixed states. These states have been used to demonstrate the phenomenon of hidden non-locality: certain Gisin states that are initially local lose this characteristic when subjected to purely local operations \cite{gisin,gs1,gs2,sbiri}. The Gisin states can be expressed using the following formula 
\begin{equation}\label{eq14}
	\mathcal{S}_{G}=\alpha | \phi_{\varphi}\rangle \langle \phi_{\varphi}| + \frac{(1 -\alpha)}{2} \left( |00\rangle\langle00| + |11\rangle\langle 11| \right),
\end{equation}
where $\alpha$ is some real number (state parameter) which satisfies $\alpha\in [0, 1]$ and the angle $\varphi \in [0, \pi/2 ]$ where $| \phi_{\varphi}\rangle = \sin(\varphi)|01\rangle + \cos(\varphi) |10\rangle$ is the Bell-like state. Gisin states are not a part of Weyl states, but they have a non-empty intersection since they are locally maximally mixed for $\varphi= \frac{\pi}{4}$. The Gisin states (\ref{eq14}) can be rewritten in the Bloch expansion as \cite{gisin,sbiri}.
\begin{widetext}
	\begin{eqnarray}\label{g2}
		\mathcal{S}_{G}=\frac{1}{4} \big[\hat{\tau}_0 \otimes\hat{\tau}_0-\alpha  \cos(2\varphi)(-\hat{\tau}_0\otimes \hat{\tau}_z + \hat{\tau}_z\otimes \hat{\tau}_0) -\alpha  \sin(2\varphi) (\hat{\tau}_x \otimes \hat{\tau}_x + \hat{\tau}_y \otimes \hat{\tau}_y) + (1-2\alpha) (\hat{\tau}_z \otimes \hat{\tau}_z)\big].
	\end{eqnarray}
\end{widetext}
To access physical resources of interest, we transform the states of ($O_2$) based on \ref{sh}. This requires us to trace over region II, which is causally disconnected from the outside. The trace over region II results in a reduced density matrix for a bipartite system consisting of a free-falling observer ($O_1$) and a static observer ($O_2$)
\begin{widetext}
	\begin{eqnarray}\label{eq15}
		\mathcal{S}_{AI} &=& \frac{(1+e^{-\vartheta})^{-1}}{4} \big[\hat{\tau}_0  \otimes \left| 0\right\rangle_I \left\langle 0\right|+(1+2e^{-\vartheta})\hat{\tau}_0  \otimes \left| 1\right\rangle_I \left\langle 1\right| - a  \cos(2\varphi)(-\hat{\tau}_z  \otimes  \left| 0\right\rangle_I \left\langle 0\right|  +(1+2e^{-\vartheta})\nonumber\\&\times& \hat{\tau}_z \otimes \left| 1\right\rangle_I \left\langle 1\right|-\hat{\tau}_0  \otimes \hat{\tau}_z) -\alpha  \sin(2\varphi)(1+e^{-\vartheta})^{1/2} (\hat{\tau}_x \otimes \hat{\tau}_x + \hat{\tau}_y \otimes \hat{\tau}_y) + (1-2\alpha) (\hat{\tau}_z \otimes \hat{\tau}_z)\big].
	\end{eqnarray}
\end{widetext}
\section{Particle detectors model  \label{sec1}}
\begin{figure}[H]
	\centering
	\includegraphics[width=0.9\linewidth]{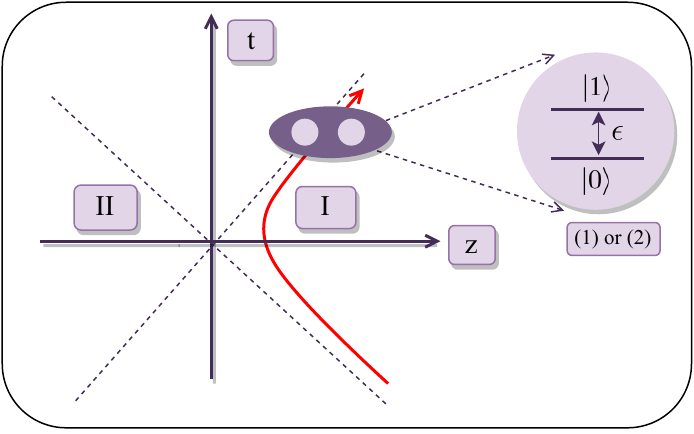}
	\caption{One-dimensional Rindler space-time diagram. Here, the curve is plotted for particle detectors with uniform acceleration. The two regions, I and II, are causally disconnected.}
	\label{fig:a}
\end{figure}

We investigate a uniformly accelerated particle detector in its vacuum state surrounded by a QFSF. 
The system can be effectively described as an open system that weakly interacts with a heat bath. In the co-moving frame as shown in Fig. \ref{fig:a}, the detector's evolution over time can be described by a master equation. This equation takes into account decoherence effects, which cause the detector's state to converge towards a purely thermal equilibrium state. This equilibrium state is distinguished by the Unruh temperature \cite{benatti2004entanglement}.
We assume that the detectors do not interact with each other and only have weak interactions with the external QFSF. In this case, the total Hamiltonian for the two detectors can be expressed as the sum of two terms
\begin{equation}
	\mathcal{H}_D=\frac{\varepsilon}{2}\sum_{j=1}^{3}r_j\Pi_{j},
\end{equation}
where $\Pi_{j}= {\hat{\tau}_{j}^{(1)}}\otimes\hat{\tau}_{0}^{(2)}+\hat{\tau}_{0}^{(1)}\otimes{\hat{\tau}_{j}^{(2)}}$ are two-system operators that are symmetric, $\varepsilon$ represents the energy level spacing of the detectors and $r_i$ is a unit vector. The Pauli matrices ${\hat{\tau}_{j}^{(1,2)}}$ refer to the detectors $(1)$ and $(2)$, while the identity matrix $\mathbb{I}^{(1,2)}$ represents the other detector. The contribution arising from the interaction of detectors with a QFSF is the combined effect of individual detector-field interactions. If the field operators $\chi_\mu$ satisfy the Klein-Gordon equation, the contribution can be expressed as follows:
\begin{equation}
	\mathcal{H}_{I}=\sum_{\mu=0}^{3}\left[  (\hat{\tau}_\mu^{(1)}\otimes \hat{\tau}_{0}^{(2)})\chi_\mu(t,\textbf{x}_1)	+(\hat{\tau}_{0}^{(1)}\otimes\hat{\tau}_\mu^{(2)})\chi_\mu(t,\textbf{x}_2)	\right].
\end{equation}
The Hamiltonian of the total system is given by
\begin{equation}
	\mathcal{H}=\mathcal{H}_D+\mathcal{H}_{\chi_\mu}+\upsilon\mathcal{H}_{I},
\end{equation}
where $\mathcal{H}_{\chi_\mu}$ is the field Hamiltonian and  expressed by means of creation and annihilation operators. However, this is not crucial for us as we will be integrating the degrees of freedom. The dimensionless coupling constant between the accelerated detectors and the field is $\upsilon\ll 1$, and the initial state of the total system is $\mathcal{S}_{tot}(0)=\mathcal{S}_{U}(0)\otimes\mathcal{S}_{F}$. Here, $\mathcal{S}_{U}(0)$ represents the initial state of the two detectors, and $\mathcal{S}_{F}$ is the similar for the entire set of QFSF considered to be stationary. The latter is given by $\mathcal{S}_{F}=|0\rangle\langle 0|$, where $|0\rangle$ is the QFSF vacuum state. At the proper time $t'$ the dynamics of the two detectors is given by the von Neumann equation $\dot{\mathcal{S}}_{tot}(t')=-i[\mathcal{H},\mathcal{S}_{tot}(t')]$. The state of the two detectors is derived by tracing over the field degrees of freedom $\mathcal{S}_{U}(t')=Tr_{\chi_\mu}(\mathcal{S}_{tot}(t'))$, and it takes the Kossakowski Lindblad form \cite{gorini1976completely,lindblad1976generators} in the limit of weak-coupling
\begin{equation}\label{deu}
	\frac{\partial \mathcal{S}_{U}(t')}{\partial t'}=-i\left[ \mathcal{H}_{eff} , \mathcal{S}_{U}(t')\right] +\mathcal{L}\left[ \mathcal{S}_{U}(t')\right],
\end{equation}
where $\mathcal{L}$ is the  Lindbladian operator given as function of Kossakowski matrix coefficients $X_{ij}$, by
\begin{equation}
	\mathcal{L}=\sum_{\substack{i,j=x,y,z \\ k,l=1,2}} \frac{X_{ij}}{2}\left[ 2\hat{\tau}_j^{(k)}\mathcal{S}_{U}\hat{\tau}_i^{(l)}-\left\lbrace \hat{\tau}_i^{(k)}\hat{\tau}_j^{(l)},\mathcal{S}_{U} \right\rbrace \right].
\end{equation}
In addition, the external quantum QFSF adds a further contribution to the detector's effective Hamiltonian, $\mathcal{H}_{eff}=\frac{1}{2}\tilde{\varepsilon}\hat{\tau}_{3}$, which takes into account the Lamb shift. This contribution can be expressed as a renormalized frequency, denoted $\tilde{\varepsilon}=\varepsilon+i\left[ \Theta(-\varepsilon)-\Theta(\varepsilon)\right]$, where the function $\Theta(y)=\frac{1}{i\pi}P\int_{-\infty}^{+\infty}d\varepsilon\frac{\Xi(\varepsilon)}{\varepsilon-y}$ is the Hilbert transform of the Wightman function. When we examine the trajectory of fast-moving detectors, it becomes clear that the Wightman function of the field satisfies the Kubo-Martin-Schwinger condition $\Gamma^+(t')=\Gamma^+(t'+i\beta_u)$, where $\beta_u=\frac{1}{T_U}$. If we transpose this principle to the frequency domain, we see that
\begin{equation}
	\Xi(y)=e^{\beta_u \varepsilon}\Xi(-y),
\end{equation}
After applying the translation invariance principle, which states that $\left\langle 0\right| \chi_\mu(x(0))\chi_\mu(x(t'))\left| 0\right\rangle =\left\langle0\right| \chi_\mu(x(-t'))\chi_\mu(x(0))\left| 0\right\rangle$, and performing specific algebraic operations, the equation (\ref{gam}) can be manipulated into a solvable expression.
\begin{equation}
	\Omega_+=\int_{-\infty}^{+\infty}\left\langle0\right|\left\lbrace \chi_\mu(t'),\chi_\mu(0)\right\rbrace\left| 0\right\rangle =(1+e^{-\beta_u\varepsilon}) \Xi(\varepsilon),
\end{equation}
\begin{equation}\label{gm}
\Omega_-=\int_{-\infty}^{+\infty}\left\langle0\right|\left\lbrace \chi_\mu(t'),\chi_\mu(0)\right\rbrace\left| 0\right\rangle =(1-e^{-\beta_u\varepsilon}) \Xi(\varepsilon),
\end{equation}
With regard to the changes caused by the interaction between the system and an external field, the next step is to define the Wightman function for a QFSF as $\Gamma^+(x,x')=\left\langle 0\right| \chi_\mu(x) \chi_\mu(x')\left|0 \right\rangle$. After that, we need to calculate its Fourier transform
\begin{eqnarray} \label{trans}
	\Xi(y)&=&\int_{-\infty}^{+\infty} dt' e^{iy t'}\Gamma^+(t')\nonumber\\&=& \int_{-\infty}^{+\infty}dt'e^{iy t'}\left\langle \chi_\mu(t')\chi_\mu(0)\right\rangle.
\end{eqnarray}
The coefficients $X_{ij}$ are calculated using the transformation outlined in (\ref{trans}). One can get
\begin{equation}\label{cof}
	X_{ij}=\Omega_+\delta_{ij}-i\Omega_- \epsilon_{ijk}r_k+\Omega_{0}r_ir_j,
\end{equation}
where
\begin{equation}\label{gam}
	\Omega_{\pm}=\Xi(\varepsilon)\pm \Xi(-\varepsilon), \quad \Omega_{0}=\Xi(0)-\frac{\Omega_+}{2}.
\end{equation}
\section{Results and analysis \label{sec4}}
Our task is to calculate and visualize the variations of quantum resources outside SBH in curved space-time. We will begin by examining a bipartite Gisin state involving a free-falling observer ($O_1$) and a static observer ($O_2$). Then, we will investigate a bipartite system consisting of uniformly accelerating particle detectors coupled with a QFSF. Our objective is to determine the quantum resources for two static detectors in a QFSF in the HH and Boulware vacuums. By doing so, we will be able to quantify the degradation of resources caused by the Hawking and Unruh effects based on specific physical parameters. These parameters include the distance between ($O_2$) and the EH $R_0$, the Gisin state parameters ($\alpha,\varphi)$, the frequency of the mode $\omega$ entangled with ($O_1$)'s field state, the energy level spacing $\varepsilon$, and the initial state choice $\kappa_{0}$.
\subsection{Hawking radiation}
Now, we calculate quantum resources for the state given in Eq. \ref{g2} by using Eq. \ref{c1} to compute quantum coherence and Eq. \ref{d1} to determine the discord. Remarkably, our analysis shows that these quantities are equivalent.
\begin{equation}\label{pl}
	\mathcal{C}_H=	\mathcal{D}_T=\left| \frac{\alpha \sin (2 \varphi )}{\sqrt{e^{-\vartheta}+1}}\right|.
\end{equation}
To assess entanglement, the first step is to compute the concurrence $\mathcal{C}$ where the values of $\lambda_{i=1,2}$ that correspond to the state Eq. \ref{eq15} can be expressed as follows:
\begin{equation}
	\lambda_{1}=-\frac{\sqrt{\alpha \cos ^2(\varphi ) e^{\vartheta} \left(-\alpha \cos (2 \varphi )+2\alpha \sin ^2(\varphi )
			e^{\vartheta}+1\right)}}{\sqrt{2}\left(e^{\vartheta}+1\right)},
\end{equation}
and
\begin{equation}
	\lambda_{2}=\frac{1}{2} \left(\left| \frac{\alpha \sin (2 \varphi )}{\sqrt{1+e^{-\vartheta}}}\right| -\frac{\sqrt{(\alpha-1)(1-\alpha\cos (2 \varphi))	e^{\vartheta}-1}}{e^{\vartheta}+1}\right).
\end{equation}
Thus, one obtain the Bures distance $\mathcal{B}_d$ which gives the quantum entanglement value for the physically accessible system $\mathcal{S}_{AI}$.

To further explore the quantum resource distribution in the physically accessible system $\mathcal{S}_{AI}$, we show the variation of $\mathcal{C}_H$, $\mathcal{D}_T$, and $\mathcal{B}_d$ in Fig. \ref{figure2} by adjusting the Hawking temperature $T_H$ and the state parameter $\alpha$, with fixed $\omega=10$, $\varphi=\pi/4$, and $R_0=1.1$.
\begin{widetext}
	\begin{minipage}{\linewidth}
		\begin{figure}[H]
			\centering
			\subfigure[]{\label{figure2a}\includegraphics[scale=0.5]{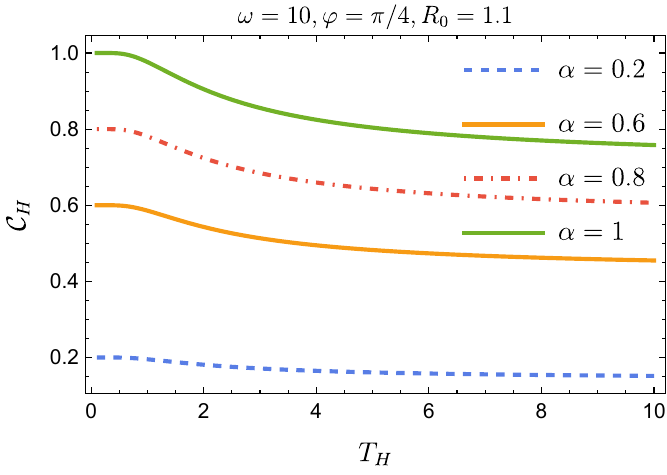}}
			\subfigure[]{\label{figure2b}\includegraphics[scale=0.5]{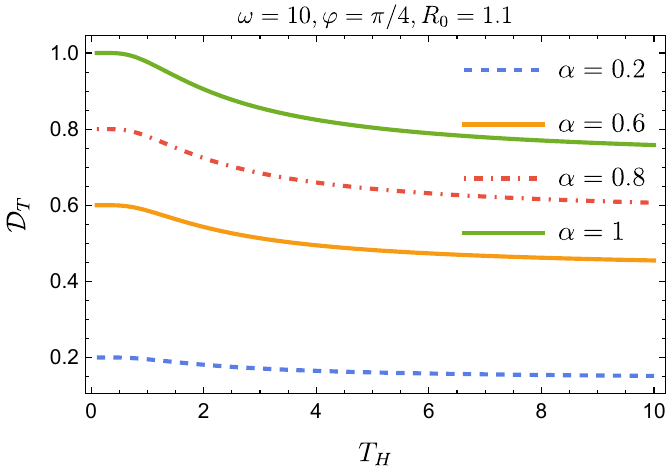}}
			\subfigure[]{\label{figure2c}\includegraphics[scale=0.5]{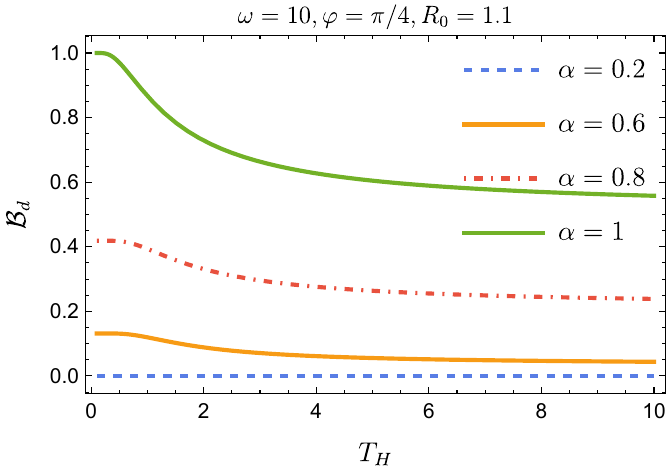}}
			\caption{Helinger distance coherence $\mathcal{C}_H$ \ref{figure2a}, trace distance discord $\mathcal{D}_T$ \ref{figure2b} and Bures distance entanglement $\mathcal{B}_d$ \ref{figure2c} for different values of the state parameter $\alpha$ against Hawking temperature $T_H$ in the physical accessible system $\mathcal{S}_{AI}$.}
			\label{figure2}
		\end{figure}
	\end{minipage}
\end{widetext}
In Fig. \ref{figure2}, a common maximum initial value, $\mathcal{C}_H=\mathcal{D}_T=\mathcal{B}_d=1$ comes into our sight near zero Hawking temperature for $\alpha=1$, and $\varphi=\pi/4$. This value is ascribed to the entangled and coherent nature of the Gisin state in the case of a supermassive SBH (i.e. $T_H=0$), where thermal Hawking radiation is absent. In addition, one can observe that the $\mathcal{C}_H$ and $\mathcal{D}_T$ curves exhibit various characters. Indeed, as the value of $\alpha$ increases from 0.2 to 1, the values of these accessible quantum resources present almost a straight line corresponding to their maximum values at first, then degrade slightly before stabilizing with the increase in Hawking temperature. These interesting phenomena can be interpreted as follows: Increasing $T_H$ can force smaller SBH to radiate more violently than massive SBH (decreasing $T_H$) and further extend the thermal decoherence generated by the Hawking effect between the fluctuating vacuum QFSF and the SBH, which causes a degradation \cite{pan2008}. In the vicinity of the SBH, fluctuating vacuum QFSF can cause indirect interplay between completely separated subsystems via correlations that enhance quantum resources \cite{hu2011entanglement}. Notably, $\mathcal{C}_H$ and $\mathcal{D}_T$ coincide with the value of $\alpha$ for the low-temperature values $0 \leqslant T_H < 1.5$. It is also peculiar that in the limit of finite Hawking temperature, the entanglement distribution undergoes distinct changes. For instance, when $T_H <1$, they remain constant, while at high temperatures, $T_H>2$, they abruptly decline, which means that the entanglement is lost due to the thermal fields generated by the thermal Hawking noise. Hence, the mixed state $\mathcal{S}_{AI}$ is always entangled for $\varphi=\pi/4$ and $0.562<\alpha\leqslant1$ in the limit $T_H \rightarrow +\infty$. Unlike $\mathcal{C}_H$ and $\mathcal{D}_T$, the value of $\mathcal{B}_d$ at $T_H = 0$ does not correspond to the value of $\alpha$. Since the entanglement in the infinite Hawking temperature limit is degraded to a higher degree than the coherence and discord, we are safe to say that the coherence and discord bound the entanglement in this limit. These results are comparable to those reported in Ref. \cite{pan2008} where the entanglement is lost for any state parameter rapidly than the mutual information due to the Hawking decoherence. However, it can be seen that the values of the state parameter $\alpha$ determine the magnitude of the quantum resources. Increasing the state parameter $\alpha$ gains more quantum resources, while increasing the Hawking temperature $T_H$ monotonically attenuates them. Moreover, quantum resources are no longer zero, whether or not the SBH is estimated to evaporate completely for the entangled Gisin state. 

Next, we will consider the impact of the Hawking temperature $T_H$ on the distribution of accessible quantum resources for specific values of the angle $\varphi$. To study this relationship, we held all other system parameters constant at $\omega=10$, $\alpha=1$, and $R_0=1.1$. 
\begin{widetext}
	\begin{minipage}{\linewidth}
		\begin{figure}[H]
			\centering
			\subfigure[]{\label{figure3a}\includegraphics[scale=0.5]{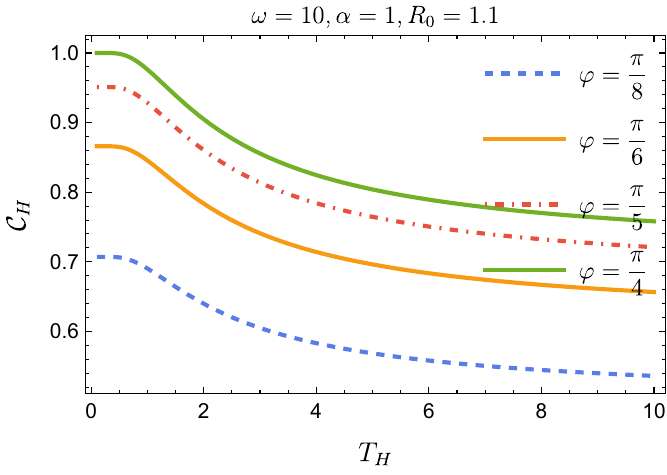}}
			\subfigure[]{\label{figure3b}\includegraphics[scale=0.5]{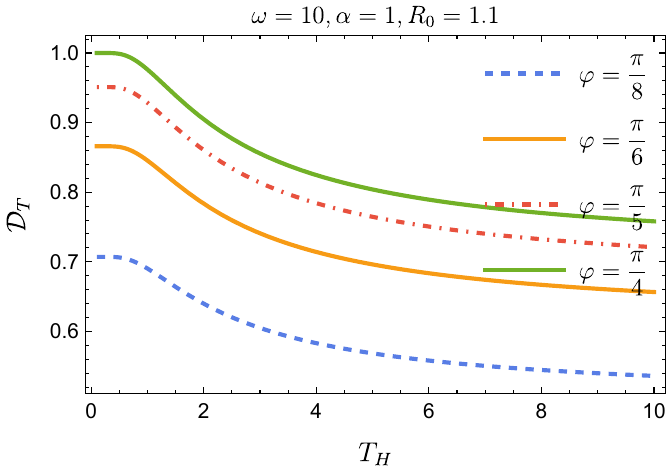}}
			\subfigure[]{\label{figure3c}\includegraphics[scale=0.5]{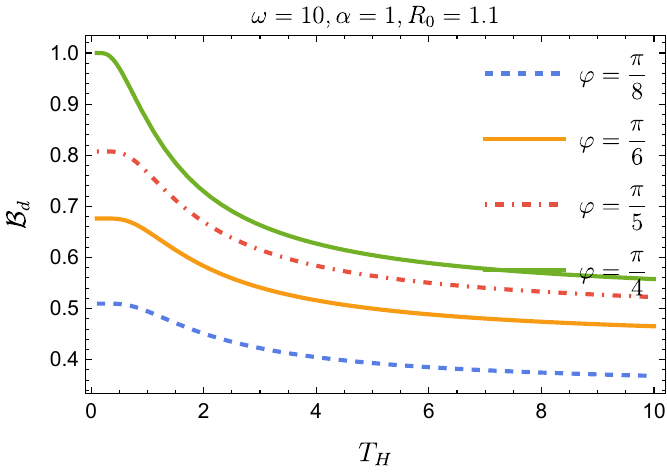}}
			\caption{Helinger distance coherence $\mathcal{C}_H$ \ref{figure1a}, trace distance discord $\mathcal{D}_T$ \ref{figure1b} and Bures distance entanglement $\mathcal{B}_d$ \ref{figure1c} against the Hawking temperature $T_H$ for different values of the angle $\varphi$ in the physical accessible system $\mathcal{S}_{AI}$.}
			\label{figure3}
		\end{figure}
	\end{minipage}
\end{widetext}

From Fig. \ref{figure3}, it is evident that the quantum resources depend on the parameter $\varphi$ as well as the Hawking temperature $T_H$, making them sensitive to the parameters of a SBH, such as mass and charge. By decreasing $T_H$, it becomes possible to achieve higher values of quantum resources. When there is no Hawking decoherence (i.e., $T_H=0$), there is a discernible generation of quantum resources, resulting in stable evolution when $T_H<2$. However, as the Hawking temperature surpasses 2 and the angle $\varphi$ decreases, there is a continuous decrease in quantum resources. For example, when $\varphi=\pi/4$ and $\alpha=1$, the curves starting from a value of 1 represent a maximally entangled state that competes with the Hawking effect. This competition leads to a significant increase in decoherence at higher temperatures, causing the degradation of resources even when optimal parameter states are chosen. At the same time, as the angle $\varphi$ decreases towards a specific value like $\varphi=\pi/8$, its influence on system resources gradually intensifies. We have discovered a remarkable result: the $\mathcal{B}_d$ is almost limited by $\mathcal{C}_H$ and $\mathcal{D}_T$. Since the initial quantum resource values vary unevenly across different $\varphi$ values, the outcomes exhibit multiple stable points around $T_H\approx10$. This is in contrast to the behavior observed in Fig. \ref{figure2}, where the state parameter of the Gisin state leads to different steady values, specifically at $T_H\approx10$. These stable values can be enhanced by carefully selecting the parameters of the Gisin state.

In order to understand the relationships between the quantum resources, the Hawking temperature $T_H$, and the mode frequency $\omega$, it is helpful to plot them in Figure \ref{figure1}. These dependencies pertain to the initial Gisin state in the physically accessible system $\mathcal{S}_{AI}$.
\begin{widetext}
	\begin{minipage}{\linewidth}
		\begin{figure}[H]
			\centering
			\subfigure[]{\label{figure1a}\includegraphics[scale=0.5]{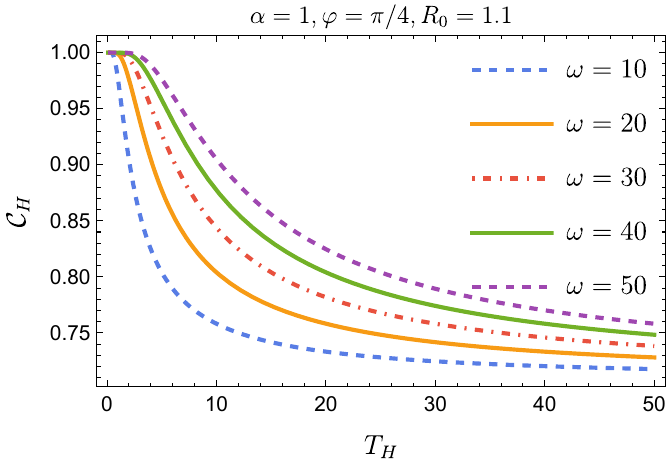}}
			\subfigure[]{\label{figure1b}\includegraphics[scale=0.5]{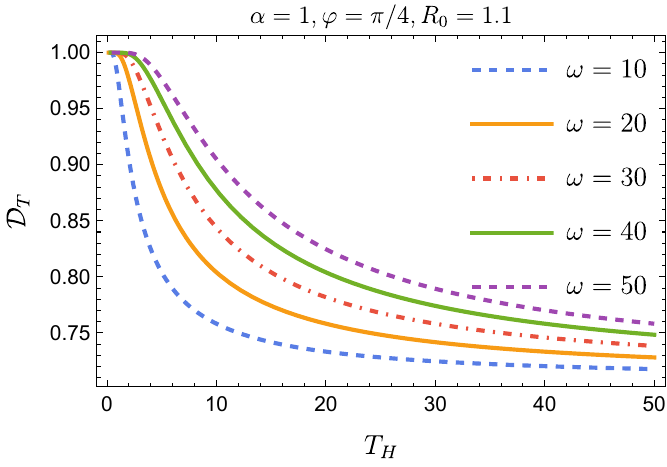}}
			\subfigure[]{\label{figure1c}\includegraphics[scale=0.5]{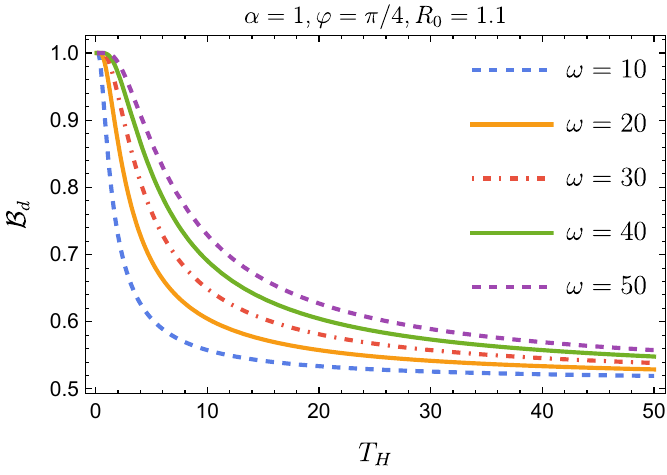}}
			\caption{Helinger distance coherence $\mathcal{C}_H$ \ref{figure1a}, trace distance discord $\mathcal{D}_T$ \ref{figure1b} and Bures distance entanglement $\mathcal{B}_d$ \ref{figure1c} for different values of the mode frequency $\omega$ in the physical accessible system $\mathcal{S}_{AI}$.}
			\label{figure1}
		\end{figure}
	\end{minipage}
\end{widetext}
The observations from Fig. \ref{figure1} reveal similarities in the profiles of the three quantum resource curves. One notable finding is that when the Hawking effect $T_H$ strengthens with a constant mode frequency parameter $\omega$, quantum resources decrease. Conversely, when the SBH temperature $T_H$ is fixed and the frequency parameter $\omega$ is adjusted, quantum resources increase as $\omega$ rises. It is worth noting that at low Hawking temperatures (approximately $T_H \approx 5$), a peak phenomenon emerges in the curves. This occurrence is attributed to the Gisin state, denoted as $\mathcal{S}_{AI}$, achieving maximum entanglement and coherence within the accessible system due to the absence of thermal radiation. Interestingly, a similar pattern is observed in the case of the Werner state \cite{shi2018quantum}. The obtained results show that in the limit of infinite Hawking temperature ($T_H\rightarrow \infty$), the physically accessible resources do not vanish and eventually reach a constant, stable value. However, higher-frequency modes can sustain resources for finite Hawking temperatures, with the mixed Gisin state being more sensitive to the increase in fermionic frequencies. It becomes evident that as the mode frequency $\omega$ decreases, all quantum resources exhibit a gradual decline. In other words, the Hawking temperature $T_H$ reduces the amount of the three quantum resources, causing them to decrease as $T_H$ increases until it reaches a steady value. On the other hand, discord and coherence limit the steady value of entanglement. This indicates that the mode frequency $\omega$ suppresses the effects of the Hawking temperature $T_H$. As the parameter $\omega$ increases, the influence of the Hawking temperature $T_H$ decreases, leading to an effective increase in the influence of the bath of QFSF outside a SBH. In other words, increasing the frequency can be a means of maintaining quantum resources.

Finally, we move on to describing the relation quantum resources, the Hawking temperature $T_H$, and the relative distance $R_0$. To gain insights, it is beneficial to represent these dependencies in Fig. \ref{figure4} for the initial Gisin state within the physically accessible system $\mathcal{S}_{AI}$.
\begin{widetext}
	\begin{minipage}{\linewidth}
		\begin{figure}[H]
			\centering
			\subfigure[]{\label{figure4a}\includegraphics[scale=0.5]{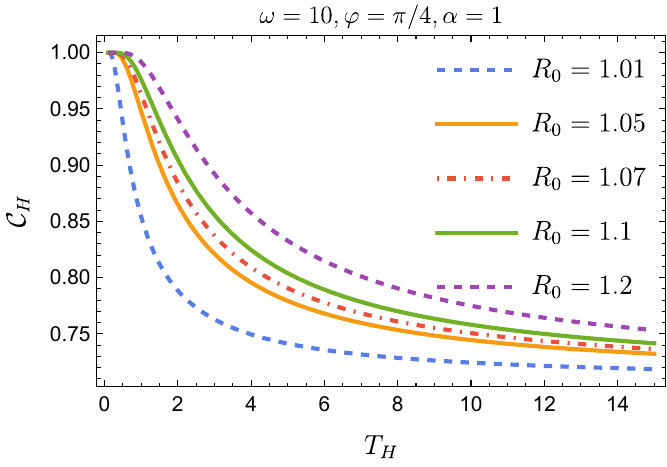}}
			\subfigure[]{\label{figure4b}\includegraphics[scale=0.5]{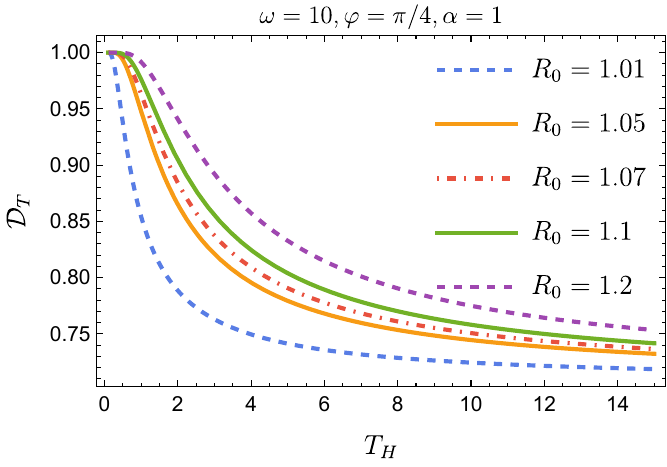}}
			\subfigure[]{\label{figure4c}\includegraphics[scale=0.5]{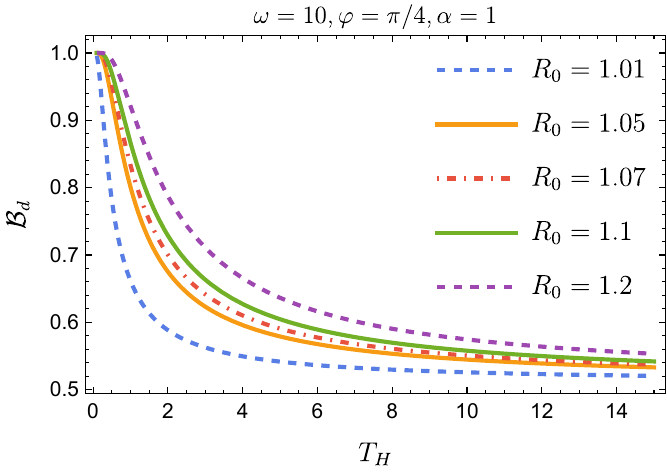}}
			\caption{Helinger distance coherence $\mathcal{C}_H$ \ref{figure1a}, trace distance discord $\mathcal{D}_T$ \ref{figure1b} and Bures distance entanglement $\mathcal{B}_d$ \ref{figure1c} for different values of the relative distance $R_0$ in the physical accessible system $\mathcal{S}_{AI}$.}
			\label{figure4}
		\end{figure}
	\end{minipage}
\end{widetext}
As shown in Fig. \ref{figure4}, in the vicinity of a SBH, we show that quantum resources can be greatly reduced to a constant value as the Hawking temperature $T_H$ increases. This implies that Hawking radiation fundamentally contributes to environmental decoherence, which in turn leads to a decrease in the quantum resources available. Remarkably, for a fixed Hawking temperature $T_H$ the quantum resources change their value several times as $R_0$ changes. At the outset, when the Hawking temperature $T_H=0$, the system exhibits maximum coherence, correlation, and entanglement. However, as the Hawking temperature exceeds 2, there is a significant decrease in quantum resources. It is worth noting that both discord and coherence follow the same behavioral pattern and converge to the same stable value, $\mathcal{D}_T\approx0.75$. On the other hand, entanglement is limited by $\mathcal{C}_H$ and $\mathcal{D}_T$ and reaches a minimum of $\mathcal{B}_d\approx0.5<0.75$. It can be concluded that the quantum resources are affected by the distance of the static observer from the EH. As both observers move farther away from the horizon, the quantum resources become more apparent. Therefore, we cannot ignore the consideration of resources in universes with EHs. Furthermore, it is enough to study changes in the vicinity of the EH, where the Rindler approximation $\frac{\Delta_H}{R_{bh}}\ll1$ is valid and distinct behaviors emerge.

\subsection{Unruh radiation}
Assuming the direction of $\vec{r} = (0, 0, 1)$, we show that the contribution of the $\Omega_{0}$ term in the Kossakowski matrix vanishes. This is what we will assume throughout the rest of the paper, unless otherwise stated. We will now demonstrate how quantum resources could be generated between the two detectors to be measured by expressing the reduced density matrix of the system in Bloch representation as follows 
\begin{eqnarray}
\mathcal{S}_{U}=\frac{1}{4}[\hat{\tau}_{0}^{(1)}\otimes\hat{\tau}_{0}^{(2)}+\sum_{j=1}^{3}s_{j}\Pi_{j}+\sum_{i,j=1}^{3}s_{ij}\hat{\tau}_{i}^{(1)}\otimes\hat{\tau}_{j}^{(2)}]. 
\end{eqnarray}
and inserting it again into (\ref{deu}), and solving a family of differential equations for a large time $t'\rightarrow\infty$, the reduced density matrix \cite{feng2015uncertainty,benatti2004entanglement} at equilibrium, is given by
\begin{eqnarray}
s_{j}=-\frac{\Omega}{3+\Omega^2}(\kappa_{0}+3)r_j,
\end{eqnarray}
and 
\begin{eqnarray}
	s_{ij}=\frac{1}{3+\Omega^2}[\Omega^2(\kappa_{0}+3)r_ir_j+(\kappa_{0}-\Omega^2)\delta_{ij}].
\end{eqnarray}
where $\Omega=\frac{\Omega_-}{\Omega_+}$ is the ratio of two constants in the Kossakowski matrix, its positivity necessitates $0\leqslant\Omega\leqslant1$. The eventual equilibrium state relies on the starting state $\kappa_{0}=\sum_{i}Tr\left[\mathcal{S}_U(0)\hat{\tau}_i^{(1)}\otimes \hat{\tau}_i^{(2)}\right]$ is a constant of motion that satisfies $\kappa_{0}\in [-3,1]$. 
This ensures that $\mathcal{S}_{U}$ remains positive. Now we can finally explicitly write down the elements of $\mathcal{S}_{U}$ as
\begin{equation}
	s_{11}=s_{44}=\frac{(3+\kappa_{0})(\Omega-1)^2}{4(3+\Omega^2)},\quad
	s_{23}=s_{32}=\frac{\kappa_{0}-\Omega^2}{2(3+\kappa^2)},
\end{equation}
\begin{equation}
	s_{22}=s_{33}=\frac{3-\kappa_{0}-(\kappa_{0}+1)\Omega^2}{4(3+\Omega^2)},
\end{equation}
with the parameter
\begin{equation}
\Omega=\frac{1-e^{-\beta_u\varepsilon}}{1+e^{-\beta_u\varepsilon}}=\tanh(\frac{\beta_u\varepsilon}{2}).
\end{equation}
The thermal nature of the Unruh effect is determined by the parameter $\beta_u$, which is dependent on the Unruh temperature $T_U$. It was proven \cite{benatti2004entanglement,feng2015uncertainty} that this equilibrium state is in general entangled if $\kappa_{0}<\frac{5\Omega^2-3}{3-\Omega^2}$.

Next, we can determine the quantum resources for the particle detectors state $\mathcal{S}_U$ by using Eq. \ref{c1} to calculate the quantum coherence and Eq. \ref{d1} to obtain the discord. Interestingly, as previously shown, it has been found that these quantities are identical.
\begin{equation}
	\mathcal{C}_H=\mathcal{D}_T=\left| \frac{\kappa _0+3}{\tanh ^2\left(\frac{\beta_u\varepsilon }{2}\right)+3}-1\right|.
\end{equation}
To proceed, note that within the context of entanglement quantification, we should first determine the concurrence $\mathcal{C}$. After a simple calculation, we get
\begin{equation}
	\lambda_{1}=\frac{\left(\kappa _0-1\right) \cosh \left(\frac{\beta_u\varepsilon }{2}\right)-2}{8 \cosh \left(\beta_u\varepsilon
		\right)+4},
\end{equation}
and
\begin{equation}
	\lambda_{2}=\frac{1}{4} \left(2 \left| \frac{\kappa _0-\tanh ^2\left(\frac{\beta_u\varepsilon }{2}\right)}{\tanh ^2\left(\frac{\beta_u\varepsilon }{2}\right)+3}\right|
	-\frac{\kappa _0+3}{2 \cosh \left(\beta_u\varepsilon \right)+1}\right).
\end{equation}
Substituting the terms above into Eq \ref{lg8}, we can calculate the Bures distance entanglement $\mathcal{B}_d$ between two particle detectors.

For a better insight of how the Unruh temperature $T_U$ and the initial state choice $\kappa_{0}$ of particle detectors impact quantum resources, we illustrate in Fig. (\ref{figure5}) the magnitudes of three quantifiers: $\mathcal{C}_H$, $\mathcal{D}_T$, and $\mathcal{B}_d$, as they vary with the Unruh temperature $T_U$ and different initial state choices $\kappa_{0}$, with $\varepsilon$ set to 5.
\begin{widetext}
	\begin{minipage}{\linewidth}
		\begin{figure}[H]
			\centering
			\subfigure[]{\label{figure5a}\includegraphics[scale=0.5]{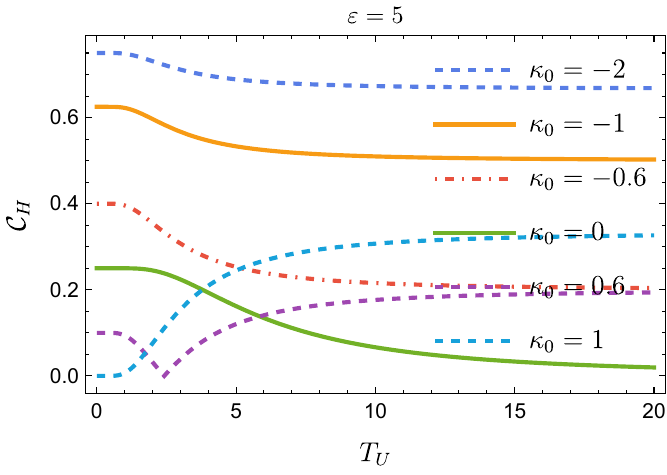}}
			\subfigure[]{\label{figure5b}\includegraphics[scale=0.5]{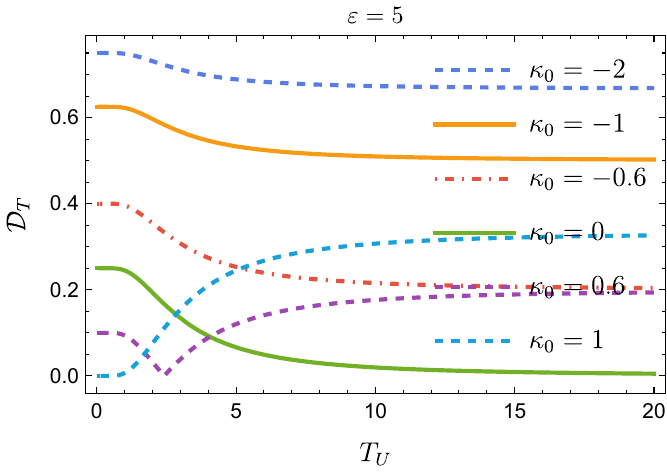}}
			\subfigure[]{\label{figure5c}\includegraphics[scale=0.5]{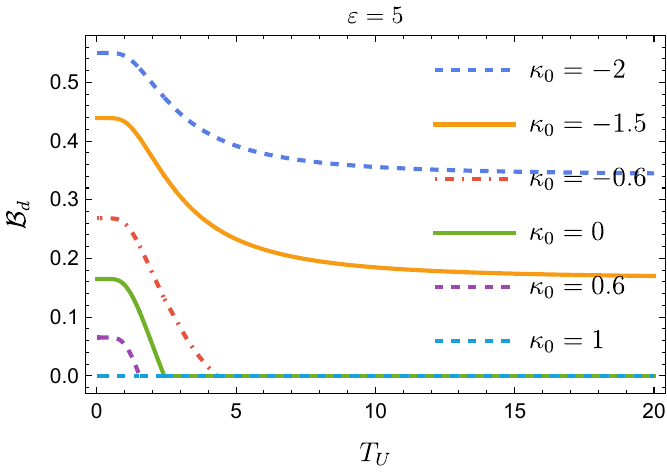}}
			\caption{Helinger distance coherence $\mathcal{C}_H$ \ref{figure5a}, trace distance discord $\mathcal{D}_T$ \ref{figure5b} and Bures distance entanglement $\mathcal{B}_d$ \ref{figure5c} for different initial state choices $\kappa_{0}$ in the particle detectors system $\mathcal{S}_{U}$.}
			\label{figure5}
		\end{figure}
	\end{minipage}
\end{widetext}
From Figs. \ref{figure5a}-\ref{figure5b}, it is clear that the $\mathcal{C}_H$ and $\mathcal{D}_T$ always decrease steadily and ultimately approach zero as the acceleration (and therefore the Unruh temperature) increases, in the case of negative initial state choices. However, there is a certain steady state at low Unruh temperatures before decay begins. This trend is due to the fact that the Unruh effect has a stronger influence on quantum resources and acts as a source of noise for this range of initial state choices. In contrast, for positive initial state choices, such as $\kappa_{0}=0.6,1$, the $\mathcal{C}_H$ and $\mathcal{D}_T$ curves show an initial decrease followed by an increase, and they finally achieve a steady value at high Unruh temperatures. In Figure 5c, we can see similar behavior in the $\mathcal{B}_d$ versus $\mathcal{C}_H$ and $\mathcal{D}_T$ curves for negative initial state values $\kappa_{0}=-2,-1,5$, except for $\kappa_{0}=-0,6$. It is noteworthy that certain choices of the initial state $\kappa_{0}$ allow detectors to generate entanglement as the Unruh temperature increases, beyond which only degradation occurs. Interestingly, when the value of $\kappa_{0}>-0.6$ is large, the $\mathcal{B}_d$ curves decrease rapidly in the low-temperature region, resulting in a decrease in the amount of entanglement and no revival of entanglement. In the case of certain negative initial state choices $\kappa_{0}<-0.6$, entanglement generation can occur without degradation. One of the reasons for this is that the degradation caused by the Unruh effect can be suppressed by choosing an appropriate initial state. It is worth noting that, in general, the Unruh-induced decoherence destroys entanglement, and as a result, the final detector state contains only classical correlations due to the presence of some correlations represented by the discord beyond entanglement.

The next step involves examining the relationship between the energy level spacing, $\varepsilon$, and the temperature of Unruh, $T_U$, to understand how it impacts the quantum resources between two moving detectors. This can be done by considering two cases: in the first situation, $\kappa_{0}$ will have a positive value of 0.1, while in the second, $\kappa_{0}$ will have a negative value of -1.5. The findings of this analysis are shown in Fig. \ref{figure7}.
\begin{widetext}
	\begin{minipage}{\linewidth}
		\begin{figure}[H]
			\centering
			\subfigure[]{\label{figure6a}\includegraphics[scale=0.5]{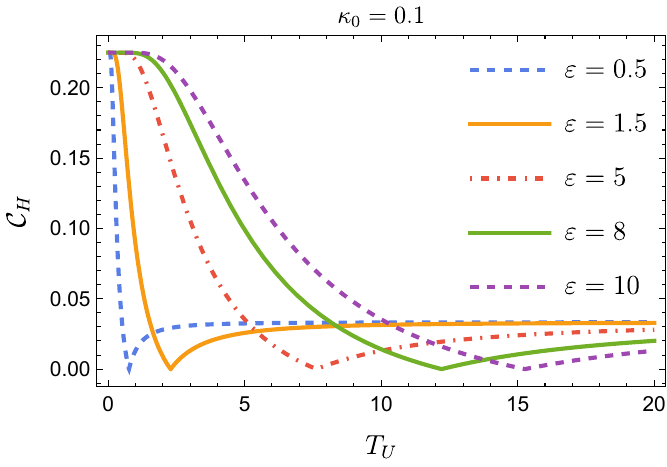}}
			\subfigure[]{\label{figure6b}\includegraphics[scale=0.5]{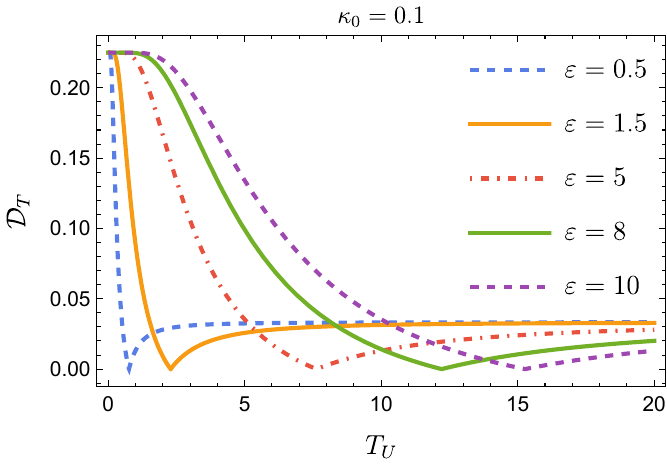}}
			\subfigure[]{\label{figure6c}\includegraphics[scale=0.5]{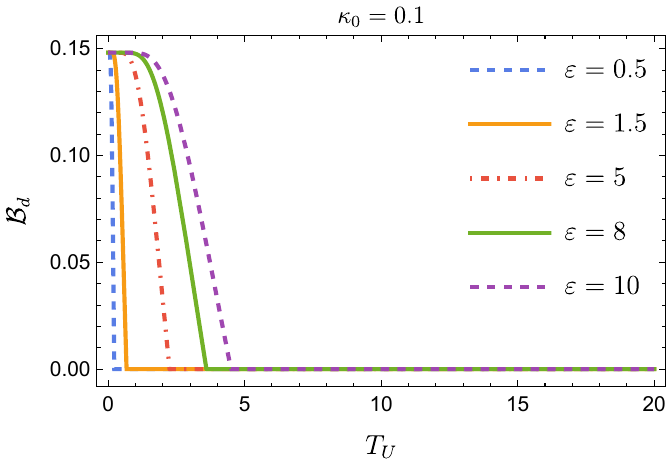}}
			\subfigure[]{\label{figure7a}\includegraphics[scale=0.5]{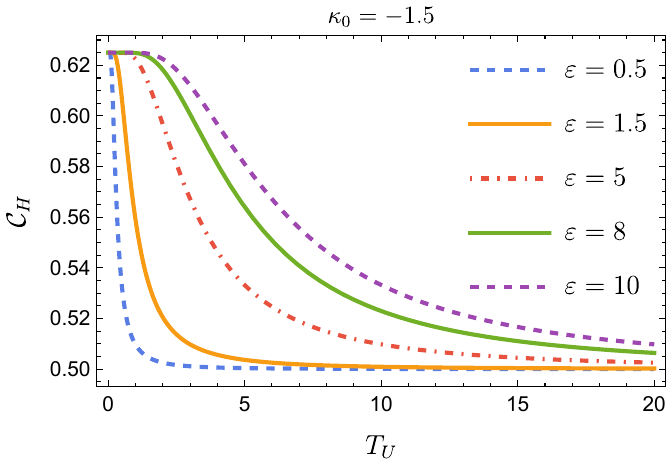}}
			\subfigure[]{\label{figure7b}\includegraphics[scale=0.5]{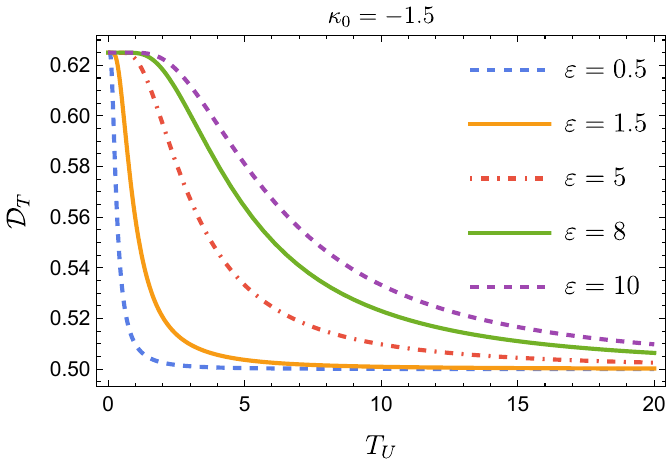}}
			\subfigure[]{\label{figure7c}\includegraphics[scale=0.5]{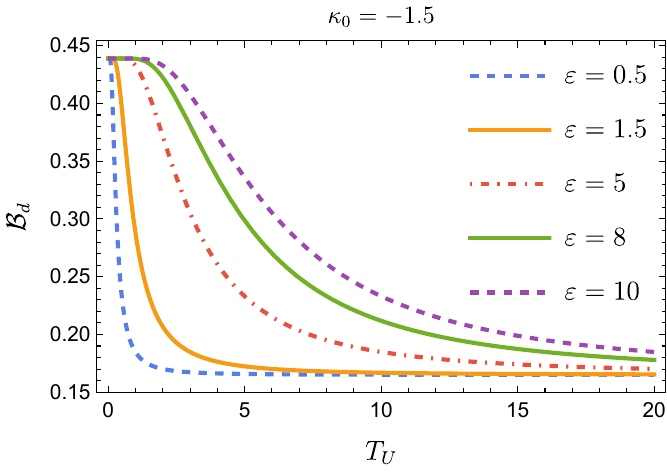}}
			\caption{Helinger distance coherence $\mathcal{C}_H$ \ref{figure1a}, trace distance discord $\mathcal{D}_T$ \ref{figure1b} and Bures distance entanglement $\mathcal{B}_d$ \ref{figure1c} for different values of the energy level spacing $\varepsilon$ in the particle detectors system $\mathcal{S}_{U}$.}
			\label{figure7}
		\end{figure}
	\end{minipage}
\end{widetext}
For a clear discussion, we first consider the case where $\kappa_0$ is positive. We find that both the $\mathcal{C}_H$ and $\mathcal{D}_T$ curves initially reach a maximum value of $0.225$ due to the absence of the Unruh effect ($T_U=0$) in each particle detector. They subsequently decline to zero at the Unruh critical temperature $T_U^c$ before steadying at roughly $0.05$. The associated Unruh critical temperatures $T_U^c$ shift to the right as the spacing of the $\varepsilon$ energy levels increases. Interestingly, for $T_U>3$, a decrease in $\varepsilon$ energy level spacing causes a rise in quantum coherence and discord. Similarly, the $\mathcal{B}_d$ curves start with a maximum value of approximately $0.15$ at $T_U=0$, so as the Unruh effect increases, we can eliminate the possibility of entanglement revival. An intriguing observation is that, when the Unruh temperature surpass the critical value, quantum coherence and discord suddenly emerge, and their degree remains stable at a certain value that depends on $\varepsilon$, independently of the increase in system temperature. Another consequence of Unruh thermal noise is the disappearance of entanglement between detectors at low Unruh temperatures, depending on the energy level spacing. These results lead us to conclude that by jointly adjusting the energy level spacing $\varepsilon$ and the Unruh temperature for a specific choice of the initial state $\kappa_0$, we can significantly enhance the quantum resources within the system. This approach offers a viable solution to reduce the negative effects of the Unruh temperature on the two-particle detector system.

We then discuss the second option, where the initial state represented by $\kappa_0$ has a negative value. Figs \ref{figure7a}-\ref{figure7c} demonstrate that we can generate quantum resources in the system under consideration, considering both the low thermal noise of the Unruh effect and the high energy level spacing of each detector. We observe a clear downward trend in the degree of $\mathcal{C}_H$, $\mathcal{D}_T$, and $\mathcal{B}_d$ as the Unruh temperature $T_U$ increases, as Unruh noise essentially degrades the resources generated between particle detectors.The curves for the coherence and discord quantifiers start from a common maximum of 0.62 and then decrease steadily to a minimum of 0.50 for higher Unruh temperatures. Moreover, the reachable value of both quantifiers falls as the $\varepsilon$ parameter is reduced. While $\mathcal{B}_d$ starts at 0.44, it then decreases to a stable value of 0.15. The results show that the quantum state is maximally coherent and correlated at $T_U=0$, and gets uncorrelated when $T_U$ exceeds 4. The results indicate that when the temperature is between $T_U<6$, we can manipulate the $\varepsilon$ parameter to enhance and sustain quantum resources in the system of particle detector pairs. However, as the Unruh temperature rises, its negative influence grows, leading to a severe degradation in quantum resources.
\subsection{Boulware vacuum state}
To fully investigate the performance of the aforementioned quantum resources on a particular curved background, we must derive the Kossakowski coefficients (\ref{cof}). These coefficients are determined by the correlation function of the QFSF in the vacuum state. In the case of SST, as mentioned in the previous section, we can identify at least two vacuum states: the Boulware vacuum state and the HH vacuum state. Let us begin with the Boulware vacuum \cite{candelas1980vacuum}
\begin{eqnarray}\label{eq32}
	\Gamma^{+}_B(x,x')&=&\sum_{jk}\int_{0}^{\infty}\frac{e^{-i\omega\Delta t} }{4\pi \omega}\left|Y_{jk}(\theta,\Phi) \right|^2\nonumber\\&\times&  \left[\left|\overrightarrow{T}_{j}(\omega,r) \right|^2+ \left|\overleftarrow{T}_{j}(\omega,r) \right|^2 \right]d\omega.
\end{eqnarray}
Whose Fourier transformation is
\begin{eqnarray}\label{eq33}
	\Xi(y)&=&\sum_{j}\frac{2l+1}{8\pi y}	\left[\left|\overrightarrow{T}_{j}(y \sqrt{f_{00}},r) \right|^2 \right.\nonumber\\&+& \left.\left|\overleftarrow{T}_{j}(y \sqrt{f_{00}},r) \right|^2 \right] \theta(y).
\end{eqnarray}
The Kossakowski coefficients may be found using $\theta(y)$ as the step function \cite{hu2011entanglement}.
\begin{equation}
\Omega_+=\Omega_-=\sum_{j=0}^{\infty}\frac{2j+1}{16\pi y}	\left[\left|\overrightarrow{T}_{j}(\omega,r) \right|^2 +\left|\overleftarrow{T}_{j}(\omega,r) \right|^2 \right],
\end{equation}
which gives $\Omega=1$.

\subsection{HH vacuum state}
The Wightman function of a QFSF for the HH vacuum has the following form \cite{candelas1980vacuum, dewitt1975quantum}:
\begin{eqnarray}\label{eq3200}
	\Gamma^{+}_H(x,x')&=&\sum_{jk}\int_{0}^{\infty}\frac{\left|Y_{jk}(\theta,\Phi) \right|^2 }{4\pi \omega}\left[ \frac{e^{-i\omega\Delta t}}{1-e^{-2\pi \omega/\varsigma}}\left|\overrightarrow{T}_{j}(\omega,r) \right|^2\right.\nonumber\\&+& \left.\frac{e^{i\omega\Delta t}}{-1+e^{2\pi \omega/\varsigma}}\left|\overleftarrow{T}_{j}(\omega,r) \right|^2 \right]d\omega.
\end{eqnarray}
Whose Fourier transformation is
\begin{eqnarray}\label{eq3220}
	\Xi(y)&=&\int_{-\infty}^{+\infty}\Gamma^{+}_H(x,x')e^{iy \tau}d\tau\\&=& \sum_{j}\frac{2j+1}{8\pi y}\left[\frac{\left|\overrightarrow{T}_{j}(y \sqrt{f_{00}},r) \right|^2}{1-e^{-2\pi y\sqrt{f_{00}}/\varsigma }}+\frac{\left|\overleftarrow{T}_{j}(y \sqrt{f_{00}},r) \right|^2}{1-e^{-2\pi y\sqrt{f_{00}}/\varsigma}} \right].\nonumber
\end{eqnarray}
Then, the Kossakowski coefficients can be explicitly given near the EH and at infinity \cite{hu2011entanglement}. By employing the geometrical optics approximation \cite{dewitt1975quantum}, we obtain a straightforward result
\begin{equation}
\Omega=\frac{e^{2\pi w_{0}\sqrt{f_{00}}/\varsigma}-1}{e^{2\pi w_{0}\sqrt{f_{00}}/\varsigma}+1}=\tanh(\vartheta).
\end{equation}
Here we use the same parameterization as in (\ref{sh}). By substituting it into (\ref{pl}), we can express the coherence and discord resources for two static detectors in a bath of the QFSF in the HH vacuum
\begin{equation}
	\mathcal{C}_H=\mathcal{D}_T=\left| \frac{\kappa _0+3}{\tanh ^2\left(\vartheta\right)+3}-1\right|.
\end{equation}
Let's first calculate the concurrence $\mathcal{C}$ between two static detectors placed in a bath of a QFSF in the HH vacuum. For the current case, we get
\begin{equation}
	\lambda_{1}=\frac{1}{4} \left(-\frac{2 \left(\kappa _0+3\right)}{\tanh ^2\left(\vartheta\right)+3}+\kappa _0+1\right),
\end{equation}
and
\begin{equation}
	\lambda_{2}=\frac{1}{4} \left(2 \left| \frac{\kappa _0-\tanh ^2\left(\vartheta\right)}{\tanh ^2\left(\vartheta\right)+3}\right| -\frac{\kappa _0+3}{2 \cosh \left(2\vartheta\right)+1}\right).
\end{equation}
Finally we get the Bures distance entanglement $\mathcal{B}_d$ for two static particle detectors.

For a deeper understanding of how changes in the positive and negative values of the initial state parameter $\kappa_{0}$ affect quantum resources, we present the values of $\mathcal{C}_H$, $\mathcal{D}_T$, and $\mathcal{B}_d$ in Fig.(\ref{figure8}), while maintaining a fixed mode frequency of $\omega=50$.
\begin{widetext}
	\begin{minipage}{\linewidth}
		\begin{figure}[H]
			\centering
			\subfigure[]{\label{figure8a}\includegraphics[scale=0.5]{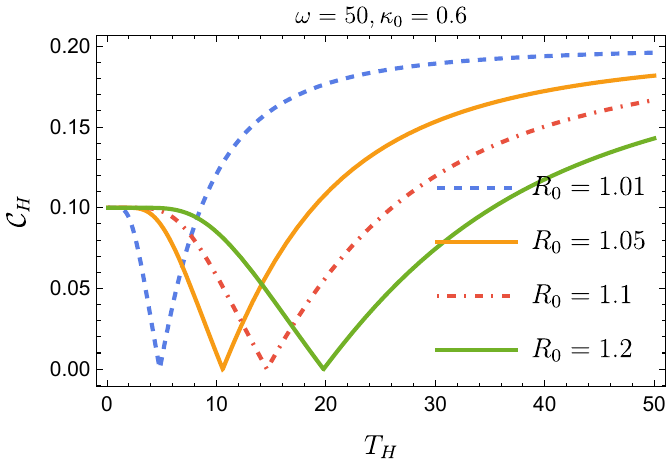}}
			\subfigure[]{\label{figure8b}\includegraphics[scale=0.5]{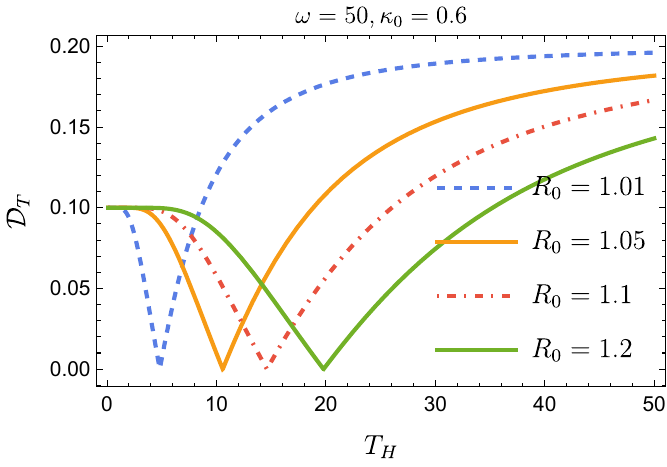}}
			\subfigure[]{\label{figure8c}\includegraphics[scale=0.5]{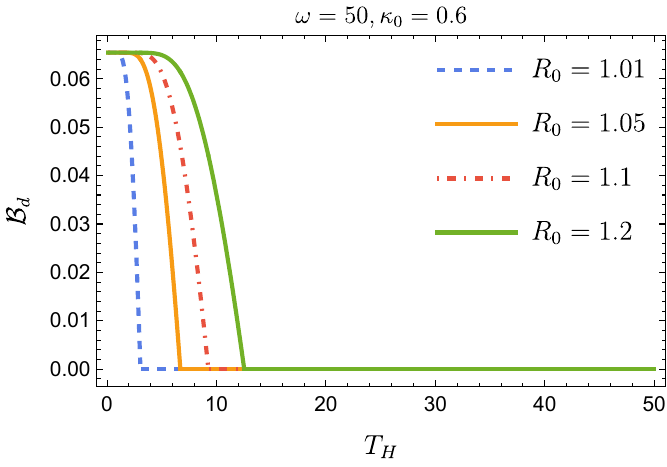}}
			\subfigure[]{\label{figure9a}\includegraphics[scale=0.5]{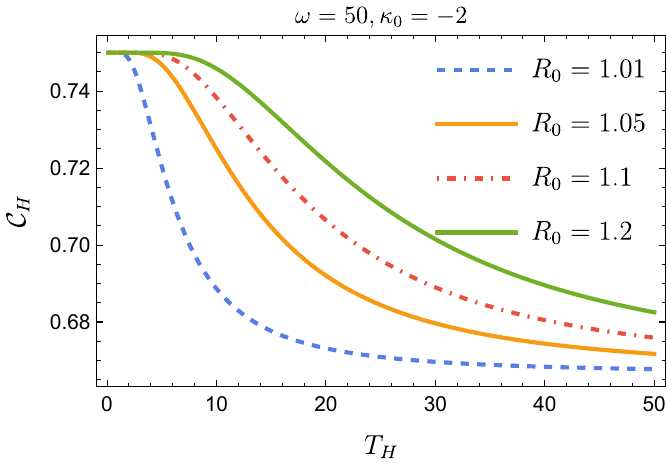}}
			\subfigure[]{\label{figure9b}\includegraphics[scale=0.5]{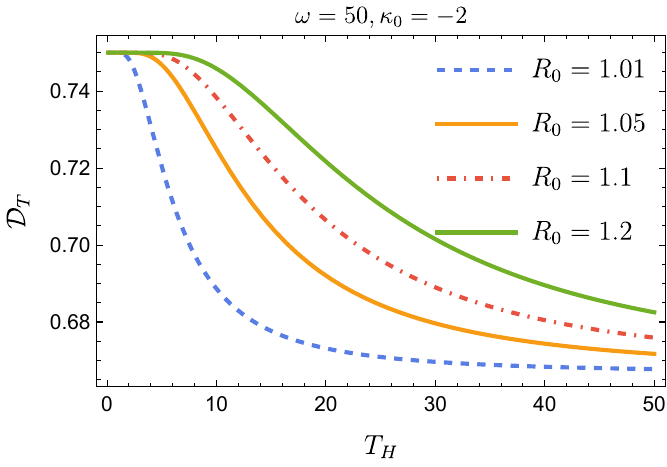}}
			\subfigure[]{\label{figure9c}\includegraphics[scale=0.5]{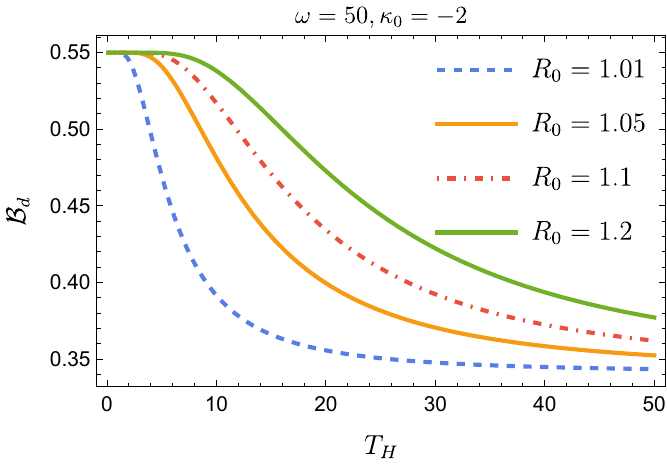}}
			\caption{Helinger distance coherence $\mathcal{C}_H$ \ref{figure8a}, trace distance discord $\mathcal{D}_T$ \ref{figure8b} and Bures distance entanglement $\mathcal{B}_d$ \ref{figure8c} for different values of the the relative distance $R_0$ in the particle detectors system $\mathcal{S}_{U}$ in HH vacuum.}
			\label{figure8}
		\end{figure}
	\end{minipage}
\end{widetext}
We begin by discussing the case of a positive initial state $\kappa_{0}=0.6$, where the quantum resources $\mathcal{C}_H$ and $\mathcal{D}_T$ are initially quite small at very low temperatures. As the temperature gradually increases, both $\mathcal{C}_H$ and $\mathcal{D}_T$ decrease until they reach the critical Hawking temperatures $T_H^c\approx 5, 10, 15, 20$, which in turn increase with the rising relative distance $R_0$. Beyond these critical temperatures, $\mathcal{C}_H$ and $\mathcal{D}_T$ begin to grow with increasing temperature, eventually surpassing the critical temperatures. Ultimately, as temperatures become significantly higher than $T_H^c$, both $\mathcal{C}_H$ and $\mathcal{D}_T$ stabilize at steady values. As shown in Fig. \ref{figure8c}, a more careful comparison reveals that the four $\mathcal{B}_d$ curves have similar profiles and start from the same maximum value $\mathcal{B}_d\approx 0.06$ at low temperatures $(T_H \leqslant 12)$. It is interesting to note that as the relative distance $R_0$ increases from 1.01 to 1.2, four sudden death values appear on the $\mathcal{B}_d$ curve. The decrease in $R_0$ causes the $\mathcal{B}_d$ curve to drop sharply in the low-temperature region, indicating a reduction in the extent of entanglement. The validity of this conclusion can be assessed by considering the fact that entanglement is sensitive to the distance of the observer's position to the EH in the HH vacuum.

Now let's focus on the second case, where the initial state has a negative value $\kappa_{0}=-2$. A notable trend emerges: as the Hawking temperature increases, there is a consistent diminishment of quantum resources. When the Hawking temperature reaches a specific threshold (where $T_H \leqslant 40$), quantum resources reach their minimum value. Further increases in the Hawking temperature lead to even more reductions in these resources, eventually stabilizing at a certain value. Similarly, an increase in $R_0$ significantly increases the quantum resources of the system at the same temperature. From a physical point of view, as both particle detectors move away from the horizon, the reduction in quantum resources becomes negligible. Consequently, quantum resources in universes containing EHs are not threatened. We also note that it is sufficient to study the change in the vicinity of the EH, where the Rindler approximation works and all characteristic behaviors can be observed. This observation aligns with specific quantitative findings in theoretical frameworks \cite{feng2015uncertainty,martin201,hu2011entanglement}.

To further explore the impact of the field mode frequency on the quantum resources in the system, we present the values of $\mathcal{C}_H$, $\mathcal{D}_T$, and $\mathcal{B}_d$ in Fig. (\ref{figure10}). We maintain a fixed relative distance of $R_0=1.1$.
\begin{widetext}
	\begin{minipage}{\linewidth}
		\begin{figure}[H]
			\centering
			\subfigure[]{\label{figure10a}\includegraphics[scale=0.5]{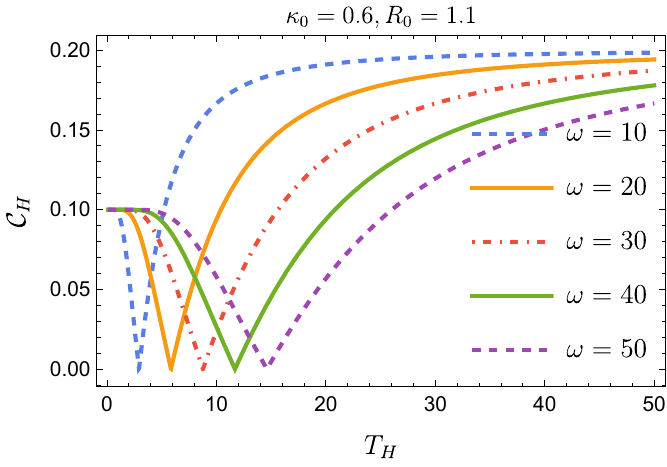}}
			\subfigure[]{\label{figure10b}\includegraphics[scale=0.5]{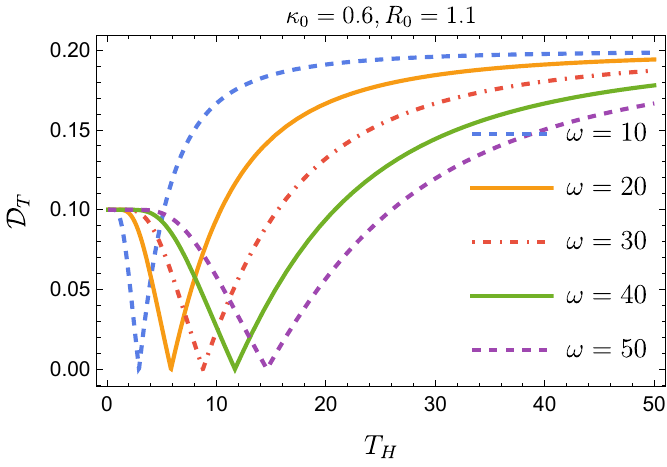}}
			\subfigure[]{\label{figure10c}\includegraphics[scale=0.5]{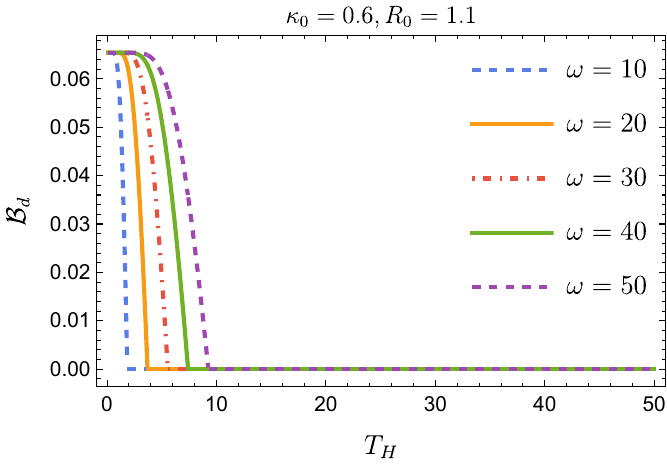}}
			\subfigure[]{\label{figure11a}\includegraphics[scale=0.5]{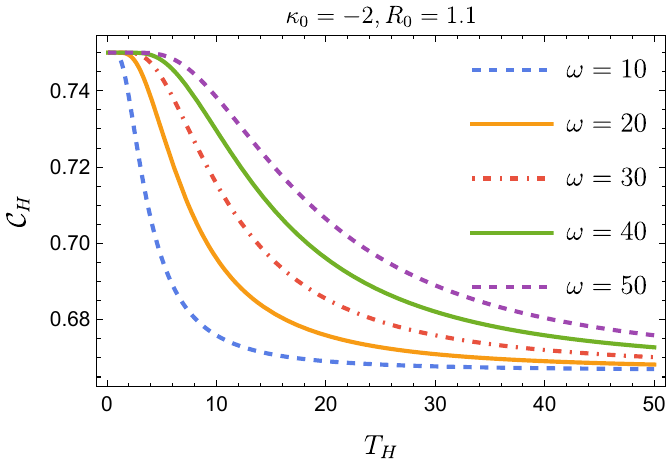}}
			\subfigure[]{\label{figure11b}\includegraphics[scale=0.5]{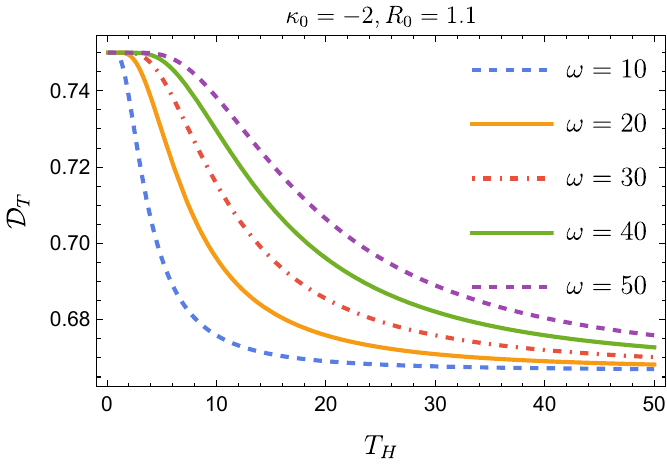}}
			\subfigure[]{\label{figure11c}\includegraphics[scale=0.5]{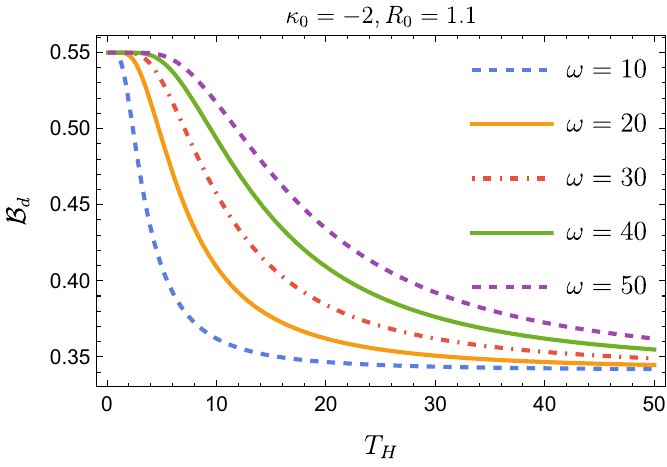}}
			\caption{Helinger distance coherence $\mathcal{C}_H$ \ref{figure10a}, trace distance discord $\mathcal{D}_T$ \ref{figure10b} and Bures distance entanglement $\mathcal{B}_d$ \ref{figure10c} for different values of the mode frequency $\omega$ for the particle detectors system $\mathcal{S}_{U}$ in HH vacuum.}
			\label{figure10}
		\end{figure}
	\end{minipage}
\end{widetext}
It is worth mentioning that the different plots in Fig. \ref{figure10} show the resulting contributions to the final density matrix of the two detectors in the HH vacuum. The first three panels show the settings with a positive initial state $\kappa_0=0.6$ for the two detectors. We can remark that the quantum resources between the detectors increase as the mode frequency of the HH vacuum decreases to $\omega=10$ below the lowest critical Hawking temperature. Additionally, the positive initial state choice induces an unusual downward trend at different mode frequencies in the variations of coherence and discord. Due to Hawking decoherence caused by increased temperatures, the entanglement between the two detectors decreases as the mode frequency decreases, eventually vanishing at a specific Hawking temperature this temperature value increases correspondingly with raising the mode frequency, as shown in Fig. \ref{figure10c}. When $T_H$ is close to zero, meaning there is no Hawking radiation, the system exhibits more entanglement. As the mode frequency decreases, the system becomes increasingly separable and incoherent. Simultaneously, we also observe that the effect of mode frequency on the amount of quantum resources is quite explicit . As the mode frequency increases, quantum resources exhibit a higher trend. In addition, the system exhibits a variety of quantum resource curves with different profiles. These profiles are influenced not only by the temperature dependence of $\mathcal{C}_H$, $\mathcal{D}_T$, and $\mathcal{B}_d$, but also by the interplay between the initial state and temperature. Furthermore, we observe a similarity between $\mathcal{C}_H$ and $\mathcal{D}_T$, where $\mathcal{C}_H$ exhibits a revival phenomenon when the initial state is positive, while entanglement rapidly disappears. On the other hand, when the initial state is negative, all resources decay monotonically towards a steady state. It is worth noting that entanglement is limited by coherence and discord resources. We can conclude that the entanglement resource appears to be very sensitive to Hawking radiation.

\section{Conclusion}\label{sec5}
In summary, we studied in depth the distribution of quantum resources from the perspective of SST using three different methods: Helinger distance coherence, trace distance discord, and Bures distance entanglement, respectively. It was shown that they provide a comprehensive understanding of the distribution of quantum resources among various subsystems in a given vacuum state. In addition, the amount of quantum resources in the physically accessible region is influenced by the Gisin state parameters, which gradually decrease due to Hawking radiation. This finding suggests that, within a specific physical environment, quantum information flows from the initial quantum system into SST as a result of Hawking radiation. For particle detectors, one can observe that as the Unruh temperature raises, quantum resources suddenly disappear, especially for certain specific initial states. However, as the Unruh effect intensifies, these resources reappear and eventually approach certain constants in the limit of infinite Unruh temperature. It is demonstrated that the Unruh effect generates resources as the energy level spacing between detectors increases, implying that the energy level spacing has a significant impact on resource generation between them. Furthermore, we find that coherence and discord are more pronounced in the presence of decoherent Unruh and Hawking noise compared to entanglement. These findings not only enhance our understanding of the dynamics of quantum resources but also highlight the vulnerability of quantum information in the face of certain physical effects. In conclusion, we believe that the results presented in our work provide fresh insights into the exploration of quantum information in the context of curved space-time, particularly from the perspective of quantum resource-based distances.

\newpage
\bibliography{references}
\bibliographystyle{ieeetr}

\end{document}